\documentclass[
preprint,
amsmath,amssymb,
aps,
pra,
]{revtex4-2}

\usepackage{graphicx}
\usepackage{dcolumn}
\usepackage{bm}
\usepackage{multirow} 

\usepackage{xcolor}
\newcommand{\note}[1]{{ \color{red} #1}}

\begin{document}


\title{{Spiraling Light: Generating Optical Tornados}}

\author{{Dimitris Mansour}$^{1,2}$}
\author{Apostolos Brimis$^{1,3}$}%
\author{Konstantinos G. Makris$^{1,3}$}
\author{Dimitris G. Papazoglou$^{1,2,}$}
\email[Corresponding author: ]{dpapa@materials.uoc.gr}

\affiliation{ $^1$Institute of Electronic Structure and Laser, Foundation for Research and Technology-Hellas (FORTH), P.O. Box 1527, 71110, Heraklion, Greece}
\affiliation{$^2$Department of Materials Science and Engineering, University of Crete, P.O. Box 2208, 70013, Heraklion, Greece}
\affiliation{ $^3$ITCP, Department of Physics, University of Crete, Heraklion, Greece}

\date{\today}

\begin{abstract}

We experimentally generate optical Tornado Waves using spatial multiplexing on a single phase modulation device. In their focal region, the intensity pattern outlines a spiral of decreasing radius and pitch. We examine the propagation dynamics of such novel waves and reveal the key factors that lead to angular acceleration. Moreover, we  propose  a  two-color  scheme that makes it possible to generate dynamically twisting light, an optical analog of a drill, that can rotate at THz frequencies. 
\end{abstract}

\maketitle

{\it Introduction.}---%
Wavefront shaping is one of the frontiers of modern photonics. Thus structured light, as we often describe the generation of customized optical fields based on wavefront shaping methods, is a topic of intense research activity due to the wide range of applications in imaging, nonlinear optics and biophotonics. Structuring of light can involve \cite{Rubinsztein-Dunlop2017}, the spatial modulation of phase, amplitude, and polarization of an optical wave. Such fields can provide a significant advantage compared to non-structured light, especially in applications where light is used as means of energy delivery on a target \cite{Papazoglou2011,Panagiotopoulos2013}. Manipulating and controlling the spatial features of the optical focus of high-power beams is a challenging problem of crucial importance for numerous applications like direct laser writing \cite{Froehly2011}, non-linear wave mixing, harmonic generation \cite{Koulouklidis2017}, and high power THz generation \cite{Liu2016}. In particular, as one increases the beam's optical power, nonlinear effects inevitably take place and as a result the beam's spatial structure is dynamically altered \cite{Couairon2007b}. 
Thus, to address such an open problem, a plethora of exotic structured optical beams have been introduced \cite{Siviloglou2007,Efremidis2010a, Papazoglou2011,Froehly2011, Chremmos2011,Rubinsztein-Dunlop2017}, beyond the well known Bessel beams \cite{Durnin1987a}. With the advent of spatial light modulation (SLM) devices it is actually possible to generate a wide range of structured light waves, optimized using various approaches. Besides direct and complex iterative numerical techniques that overshadow the physical intuitive understanding, alternative semi-analytical solutions have been recently emerged that allow us to precisely control the focal distribution. A key point of this approach is the use of non-diffracting light like Bessel beams \cite{Durnin1987a} and accelerating Airy beams \cite{Siviloglou2007}. The propagation of these optical waves is dominated by a strong linear energy flux that through interference generates the high intensity features of the beam. As the power is increased, nonlinear effects are substantial only at the high intensity regions \cite{Panagiotopoulos2013}, thus making it possible to control the propagation dynamics.
In this context, an ideal platform that enables such tailored control is the recently introduced family of rotationally symmetric accelerating beams, often referred as abruptly autofocusing beams or circular Airy beams (CABs) or ring-Airy beams, whose radial distribution is described by the Airy function \cite{Efremidis2010a, Papazoglou2011}. These waves propagate in curved trajectories, and exhibit abrupt autofocus, while at high intensities they reshape into nonlinear intense light-bullets with extremely well defined focal position \cite{Panagiotopoulos2013}.

In fact, imprinting a helical phase can induce topological charge to the wave field, and thus create an optical vortex. Such waves carry orbital angular momentum (OAM) and exhibit a rotating phase structure as they propagate \cite{Rubinsztein-Dunlop2017}. Twisting structured light, where the intensity pattern rotates forming a helical pattern, can be generated by superimposing structured light that caries {OAM}  of opposite helicity \cite{ Carmon2001,Froehly2011,Rubinsztein-Dunlop2017}. Furthermore, by properly tuning the interfering {OAM} beams, angular acceleration or deceleration can be achieved. The field thus forms a helix of variable pitch \note{uopn} propagation \cite{Schulze2015, Webster2017}. Recently, Tornado waves (ToWs), a new type of structured light that combines the radial acceleration with the angular acceleration was theoretically introduced \cite{brimis2020tornado} and preliminary experimental generation was demonstrated \cite{Mansour2021}. Like a tornado, ToWs intensity maxima outline a spiral of decreasing radius and pitch as they propagate. The combination of angular acceleration with intense abrupt autofocusing makes these novel waves ideal for applications.

In this work we study the propagation dynamics of twisting light that spirals like a tornado. Using spatial multiplexing we experimentally realize, Tornado Waves and study their property to twist and accelerate both in the radial and angular direction. Furthermore, we analyze the basic factors that lead to the effect of acceleration and it's  dependence on the number of high intensity lobes. Likewise, we propose a two-color scheme that makes it possible to generate dynamically twisting light that can rotate at $\rm{THz}$ frequencies.

{\it Twisting light and angular acceleration}---%
\begin{figure}[t]
\centering
\includegraphics[width=0.7\textwidth]{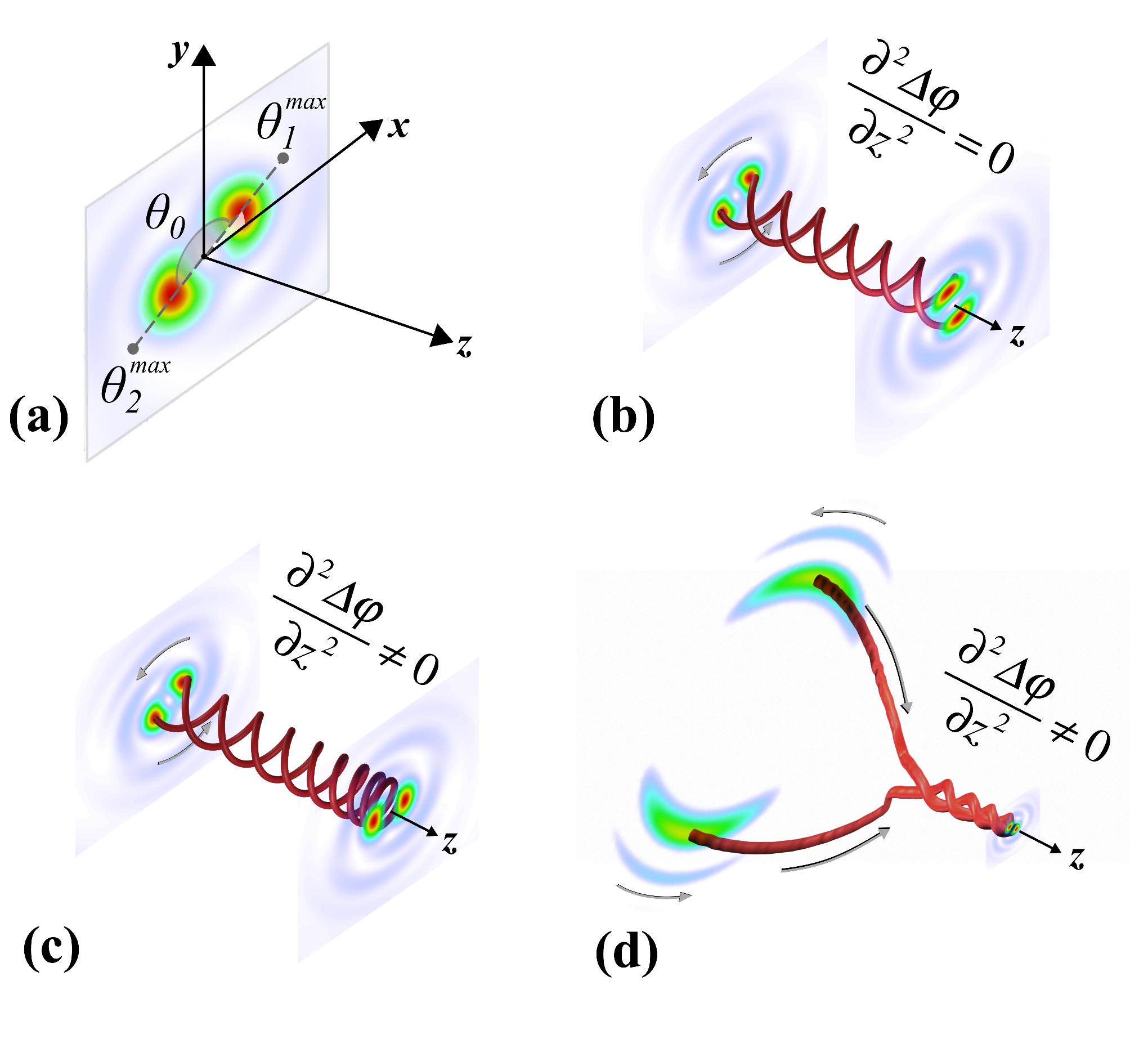}
\caption{Twisting light structures based on interference of two waves that carry OAM of opposite helicity for $t={\it const}$. (a) distribution of high intensity lobes on the transverse plane, (b) $\partial^2 \Delta \varphi/\partial z^2 =0$: constant angular velocity. (c) $\partial^2 \Delta \varphi/\partial z^2 \ne 0$: angular acceleration, and (d) radial and angular acceleration of ToWs.}
\label{fig:Acceleration_Principle}
\end{figure}

The physical origin behind the formation of a twisting light structure and it's properties can be  revealed by studying of the interference of waves that carry {OAM} of opposite helicity. The propagation of pulsed scalar beams in dispersive media is described, using a moving reference frame that follows the pulse, by the paraxial wave propagation equation:    \cite{Brabec1997,Couairon2007b}:
\begin{gather}
    2 i k \frac{\partial {u}}{\partial z} - k D_{\omega_0} \frac{\partial^2 {u}}{\partial \tau^2}+\nabla_{\bot}^2 u  = 0
\label{eq:paraxial}    
\end{gather}
where $\tau \equiv t-z/\upsilon_g(\omega_0)$ is a reduced time, $\upsilon_g(\omega_0)$ is the group velocity and  $D_{\omega_0} \equiv \partial^2 {k}/\partial {\omega}^2|_{\omega_0} $ is the group velocity dispersion (GVD) at a central frequency $\omega_0$, $\nabla_{\bot}^2$ denotes the transverse part of the Laplacian, $k$ is the wavenumber,  $z$ is the propagation distance,  $t$ is time and $u$ the electric field envelope. 

For simplicity let us consider the case of two interfering continuous waves (CW) of different frequencies $(\omega_A,~\omega_B)$. Their superposition can be described as: 
\begin{equation}
    \label{eq:TwistingLight}
    \begin{aligned}
     {u_{s}}({\bf{r}},t) = &\left| {u_A}({\bf{r}})\right|{e^{i[{\varphi _A}\left( {\bf{r}}\right) + l_A \theta - {\omega _A}t]}} \\
    + &\left| {u_B}({\bf{r}})\right|{e^{i[{\varphi _B}\left( {{\bf{r}}} \right) + l_B \theta - {\omega _B}t]}}
    \end{aligned}
\end{equation}
%
%
where $\varphi_i\left( {\bf{r}} \right)$ refers to the spatial phase profile, $\theta$ is the azimuthal angle and $l_A, l_B$ is respectively the topological charge of each wave. It  can be directly shown that  such a superposition will lead to an intensity profile $I \propto \left | u_s\right|^2$, that is described by:
\begin{equation}
    \label{eq:IntensityTwisting}
    \begin{aligned}
    I \propto & {~2\left| {{u_A}({\mathbf{r}})} \right|\left| {{u_B}({\mathbf{r}})} \right|\cos \left[ {\Delta \varphi ({\mathbf{r}}) -(l_A-l_B)~\theta  - \Delta \omega t} \right]} \\
    & {+\left| {{u_A}({\mathbf{r}})} \right|^2} + {\left| {{u_B}({\mathbf{r}})} \right|^2} 
    \end{aligned}
\end{equation}
where $\Delta \varphi ({\mathbf{r}}) = {\varphi _B}\left( {{\bf{r}}} \right)-{\varphi _A}\left( {{\bf{r}}} \right)$ is the spatial phase difference, and  ${\Delta \omega =\omega _B-\omega _A}$ is the frequency difference. Clearly, the cosine term in Eq. (\ref{eq:IntensityTwisting}) leads to an intensity modulation in the transverse plane.{We can analytically describe the evolution of this intensity structure when the spatial phase difference is not a function of the azimuth angle $\theta$:  $\Delta \varphi ({\mathbf{r}}) \equiv \Delta \varphi ({\rho,z})$, where $\rho$ is the radial distance from the $z$ axis. In this context, it is straightforward to show that an evenly distributed number of $N=\left | {l_A} - {l_B} \right |$  high intensity lobes appear at azimuthal angles $\theta^{max}_m$ described by}:
\begin{equation}
    \theta^{max}_m=\frac{1}{N}\left[ {\Delta \varphi({\rho,z}})  - \Delta \omega t \right ]+(m-1)\theta_0
        \label{eq:TwistingTheta}
\end{equation}
where $\theta_0 \equiv 2 \pi/N$, and $m=1,2 \ldots N$ is an index. {Likewise, due to the opposite handedness of the two beams $({l_A} {l_B}<0)$, the number of lobes $N$ can be rewritten as $N=\left | {l_A} \right |+\left | {l_B} \right |$}. If we freeze time $(t={\it const})$, these high intensity lobes rotate as we move along the $z$ axis and form a helical pattern as shown in Fig. \ref{fig:Acceleration_Principle}. The spatial rate of this rotation is related to the pitch of the helix and is referred to as angular velocity \cite{Schulze2015,Webster2017,brimis2020tornado}. Using  Eq.(\ref{eq:TwistingTheta}) this can be described as:
\begin{equation}
    \upsilon  =\frac{{\partial \theta^{max}_m }}{{\partial z}}= \frac{1}{N}\frac{{\partial\Delta \varphi }}{{\partial z}}
    \label{eq:scaleangvel}    
\end{equation}
\begin{figure}[b]
\centering
\includegraphics[width=0.7\textwidth]{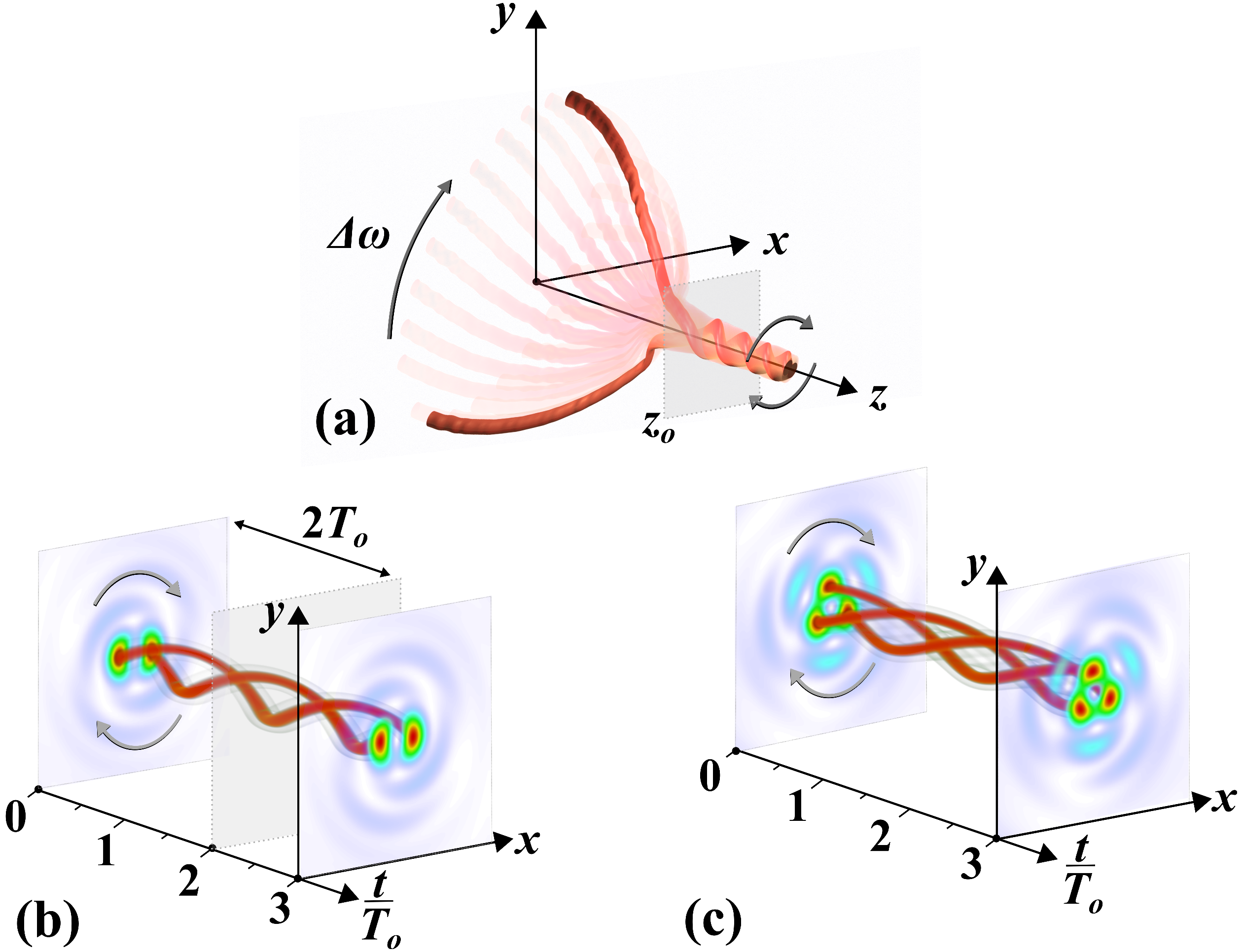}
\caption{Rapid rotation of a two color Tornado wave. (a) Graphical representation of a rapidly rotating two color tornado wave. (b),(c) Numerical simulation results depicting the  temporal evolution of the twisting lobes at $z_0$ for $N=2$ and $N=3$ high intensity lobes respectively. $(\Delta \omega \simeq 100~\rm{THz}, T_o \simeq 61~\rm{fs})$ }
\label{fig:ToWBeat}
\end{figure}
Thus, the angular velocity $\upsilon$ is related to the rate of change along $z$ of the spatial phase difference $\Delta \varphi$, and is decreasing when we increase {the number of high intensity lobes N}. Likewise, when $\Delta \varphi$ is a linear function of $z$, the superposition leads to a constant angular velocity thus, as shown in Fig.\ref{fig:Acceleration_Principle}(a), the spiral has a constant pitch. For example, we observe this behaviour when two Bessel beams, that carry {OAM} of opposite helicity, interfere \cite{Schulze2015}. In this case, it is straightforward to show that $\Delta \varphi= (k^B_z-k^A_z)z$, where $k^A_z,k^B_z$ are respectively the wavevector projections along the $z$-axis. 

On the other hand, when $\Delta \varphi$ is a non-linear function of the propagation distance $z$ the spiral has a varying pitch, as shown in Fig. \ref{fig:Acceleration_Principle}(b). The spatial rate of this variation is referred to as angular acceleration \cite{Schulze2015,Webster2017,brimis2020tornado} and by using  Eq.(\ref{eq:TwistingTheta}) it can be described as:
\begin{equation}
   \gamma \equiv \frac{\partial \upsilon}{\partial z} =\frac{{\partial^2 \theta^{max}_m }}{{\partial {z^2}}}= \frac{1}{N}\frac{{{\partial^2}\Delta \varphi }}{{\partial{z^2}}}
    \label{eq:scaleAcc}    
\end{equation}

Angular acceleration can be achieved following different approaches. One is to engineer a variable $\Delta \varphi$ starting from simple conical waves that carry OAM. For example, Schulze et al. \cite{Schulze2015} have demonstrated that if instead of a pair, two pairs of Bessel beams are used  $\Delta \varphi$ becomes a periodic function of $z$ and angular acceleration is observed. On the other hand, angular acceleration can be observed by using interfering pairs of more complex waves with an inherent non-linear $\Delta \varphi(z)$ variation, like for example Laguerre–Gaussian modes \cite{Kotlyar1997,Kotlyar2007,Webster2017}. 

Going a step further, we propose here the interferecne of two ring Airy-beams. In particular, angular acceleration is combined with radial acceleration in the case of ToWs where, as shown schematically in Fig. \ref{fig:Acceleration_Principle}(d), two ring-Airy beams carrying OAM of opposite helicity interfere to form a spiral intensity structure that twists and shrinks in an accelerating fashion \cite{brimis2020tornado}. The inherent property of radial acceleration of ring-Airy beams \cite{Efremidis2010a,Papazoglou2011,Panagiotopoulos2013} allows us to localize the angular acceleration in the abrupt autofocus area.    

In the case where $\omega_A=\omega_B$, the interfering waves have the same frequency and the twisting structures that are depicted in Fig. \ref{fig:Acceleration_Principle} are static in time.  On the other hand, if $\omega_A \ne \omega_B$ these field profiles will dynamically rotate around $z$ axis with a temporal period:
\begin{equation}
    {T_r}=N~T_o = \frac{N}{c}\frac{{\lambda _A^2}}{{\Delta \lambda }}\left( {1 + \frac{{\Delta \lambda }}{{{\lambda _A}}}} \right)
\label{eq:ToW_Tr}
\end{equation}
where $T_o = 2\pi (\omega_A-\omega_B)^{-1}$ is a reference beat period \cite{Odoulov2015}, and $\Delta \lambda = \lambda_B-\lambda_A$, where $\lambda_A$, and $\lambda_B$ {are vacuum wavelengths, respectively}. The period for a full rotation of the $N$ high intensity lobe pattern is proportional to the number of lobes. In principle, due to symmetry these high intensity lobes are identical and they are angularly distributed at $2\pi/N$ intervals, so as they rotate they will periodically overlap at a period $T_o$. 
Although this behaviour is similar to the well known effect of running fringes \cite{Odoulov2015} the symmetry is different since the high intensity lobes are rapidly rotating around a point on the transverse $(x,y)$ plane and the whole structure rotates around $z$ axis making an optical analog of a drill. Such dynamically twisting light structures are illustrated in Fig. \ref{fig:ToWBeat}. Using numerical estimations of Eq.(\ref{eq:TwistingLight}) for the case of two-color ToWs where $\lambda_A= 594~\rm{nm},~\Delta \lambda = 20~\rm{nm}$ we have visualized the temporal evolution of a two-color ToW at the autofocus position for the case of two  $(N=2, ~ l_{A,B}= \pm 1)$ and three $(N=3, ~l_{A,B}=+1,-2)$ high intensity lobes. The reference period {in this case is $T_o \sim 61~\rm{fs}$}. As  predicted by Eq.(\ref{eq:ToW_Tr}) the high intensity lobes perform a full rotation in $2 T_o \sim 122~\rm{fs}$ for $N=2$ and in $3T_o \sim 182~\rm{fs}$ for $N=3$. Such structured light, that rotates rapidly within the pulse duration {\cite{Rego2019,Zhao2020,Bejot2021,Wang2021}}, can be quite appealing in  direct laser writing applications \cite{Froehly2011}.
\begin{figure}[t]
\centering
\includegraphics[width=0.7\textwidth]{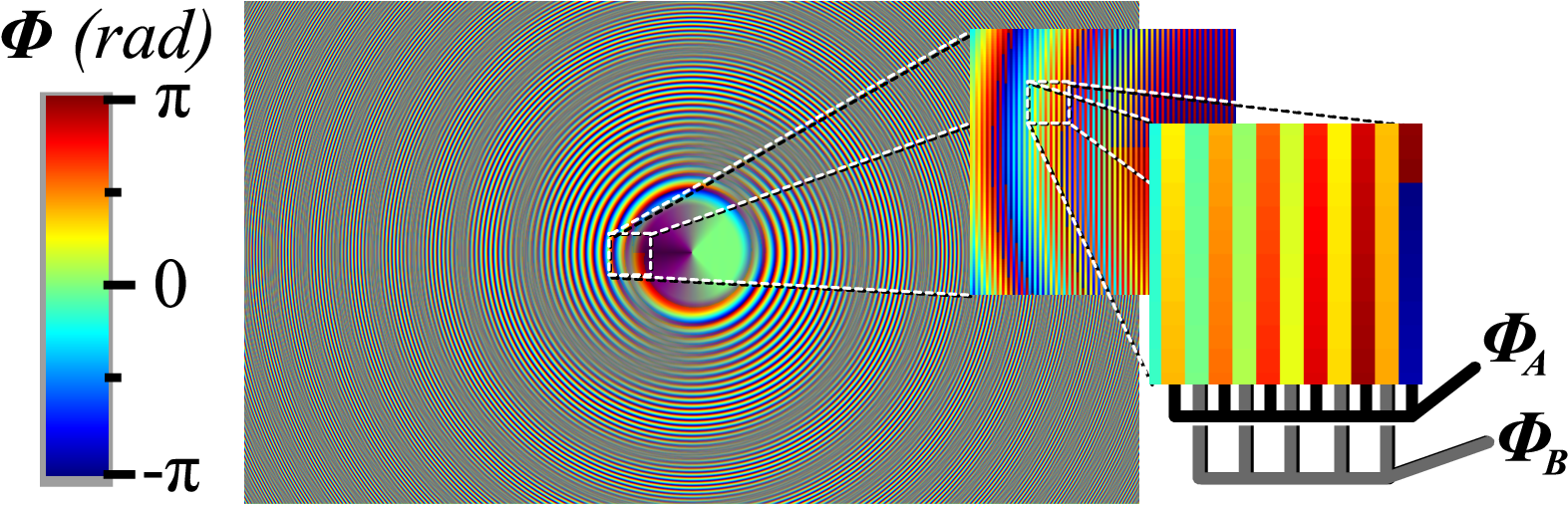}
\caption{Phase mask for the generation of ToWs {in the case of $l_A=+1,~l_B=-1$, with design parameters provided in Table \ref{tab:MaskParameters}}. Insets {show small portions of the mask that} reveal the alternating sampling geometry for the individual masks $\varPhi_A,~\varPhi_B$ {which} respectively generate each of the superimposing accelerating waves.}
\label{fig:PhaseMask}
\end{figure}
\begin{figure}[b]
\centering
\includegraphics[width=0.7\textwidth]{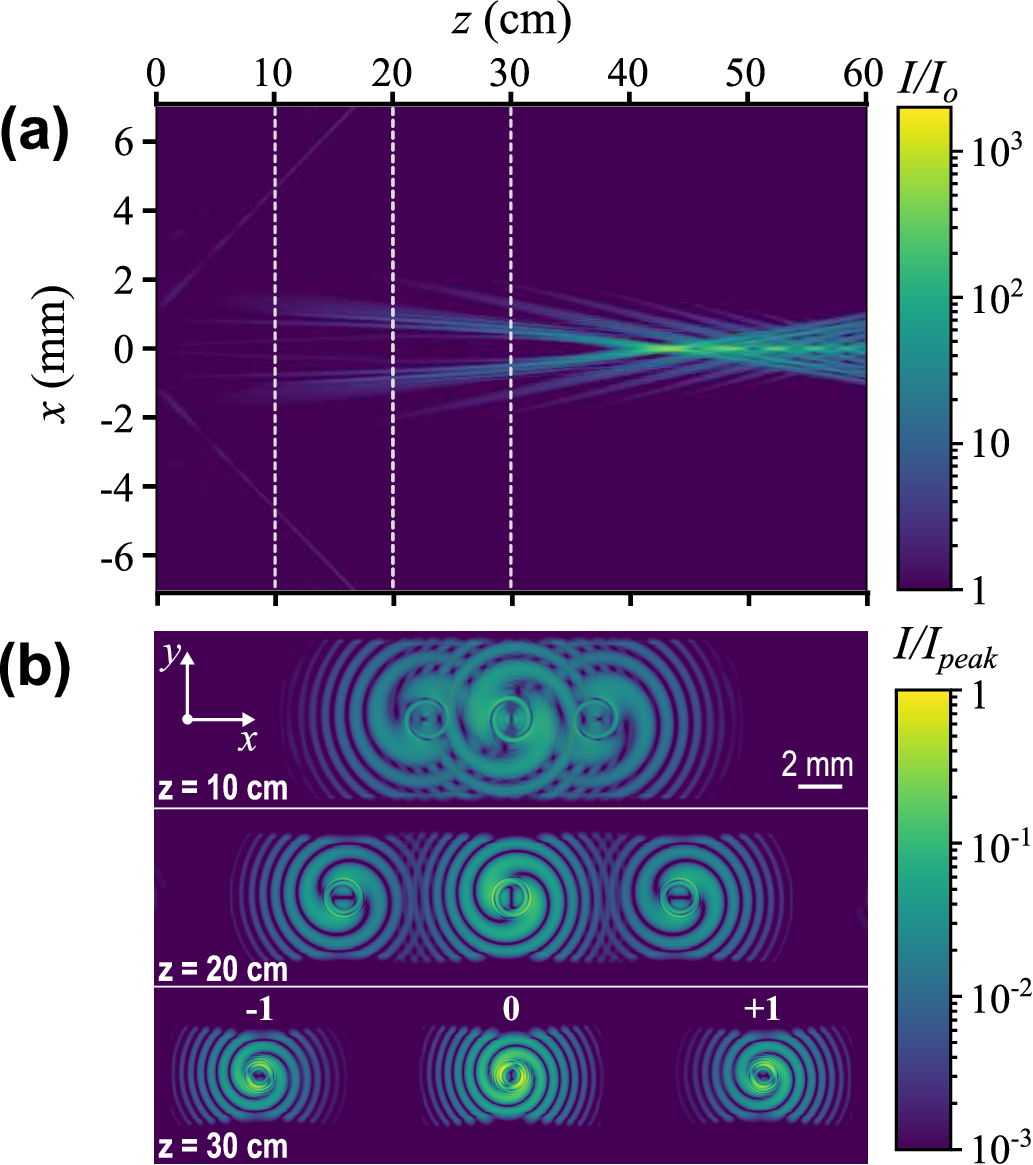}
\caption{Numerical simulation of the propagation of Tornado waves generated by the phase mask of Fig. \ref{fig:PhaseMask}. (a) Intensity profile distribution $I(x,z)$ {(normalized to the  initial peak intensity $I_o \equiv max\{I(x,y,0)\}$)} and (b) Transverse intensity profiles $I(x,y)$  at $z=10,~20,~30~$cm (normalized to {the peak intensity $I_{peak} \equiv max\{I(x,y;z)\}$} for each $z$). Note that after $30~$cm of propagation the $\pm 1$ diffraction orders are clearly separated from the zero order.}
\label{fig:I_z}
\end{figure}

{\it Realization of Tornado waves}---%
As theoretically described by Brimis {et al.} \cite{brimis2020tornado}, ToWs can be generated by superimposing two ring-Airy beams that are tuned to abruptly auto-focus at overlapping focal regions, while carrying OAM of opposite helicity \cite{brimis2020tornado}. The generation of ring-Airy beams involves the use of a phase or amplitude SLMs \cite{Papazoglou2011,Liu2016}. Phase modulation is usually preferred \cite{Siviloglou2007, Papazoglou2011}, since compared to amplitude modulation, the power efficiency can be higher \cite{Goodman1996}. On the other hand, direct generalization of phase only modulation approaches to interfering fields that carry OAM is non-trivial. For example, using a description similar to that of Eq.(\ref{eq:TwistingLight}), let's consider the simple case of two plane waves of equal amplitude and frequency $\left| {u_A}({\bf{r}})\right|=\left| {u_B}({\bf{r}})\right| \equiv 1, \omega_A=\omega_B$, {which are phase modulated to imprint on them OAM of} topological charge $l_A, l_B$ respectively. Although both of them can be independently generated using a phase SLM their superposition involves an amplitude and phase modulation $u_s({\bf{r}})=2\cos (\frac{{{l_A} - {l_B}}}{2}){e^{i\frac{{{l_A} + {l_B}}}{2}}}$ that a phase SLM cannot directly reproduce. Simply adding the two phase distributions reproduces only the phase distribution (ignoring a factor 2) of their superposition and not the amplitude. Furthermore, when the beams carry OAM of opposite helicity ${l_A=-l_B}$ this phase sum is zero. 
Here we follow an approach that involves spatial multiplexing on a single SLM device and allows superimposed fields, of any complexity in their phase structure, to be generated \cite{Arrizon2007, Mendoza-Yero2014,LuisMartinezFuentes2018, Wang2020, Davis2021, Mansour2021}.
The spatial sampling effectively acts as a diffraction grating, where the original distribution is replicated in all diffraction orders that propagate in different directions making an angle $\beta_j=sin^{-1}(\frac{j\lambda}{2 w})$ to the $z$ axis, where $w$ is the sampling period and $j=0, \pm 1 \ldots$ is the diffraction order. The angular separation $\beta_j$ is a key element of this approach since by simple propagation all diffraction orders become spatially separated and can be easily isolated.
For example for an SLM device with a typical pixel size of $w=10~\mu$m operating at $\lambda = 1~\mu$m the angular separation between the zero and the first order is $\sim 2.86^o$ corresponding to a numerical aperture of $NA \sim 0.05$.
\begin{figure}[t]
\centering
\includegraphics[width=0.7\textwidth]{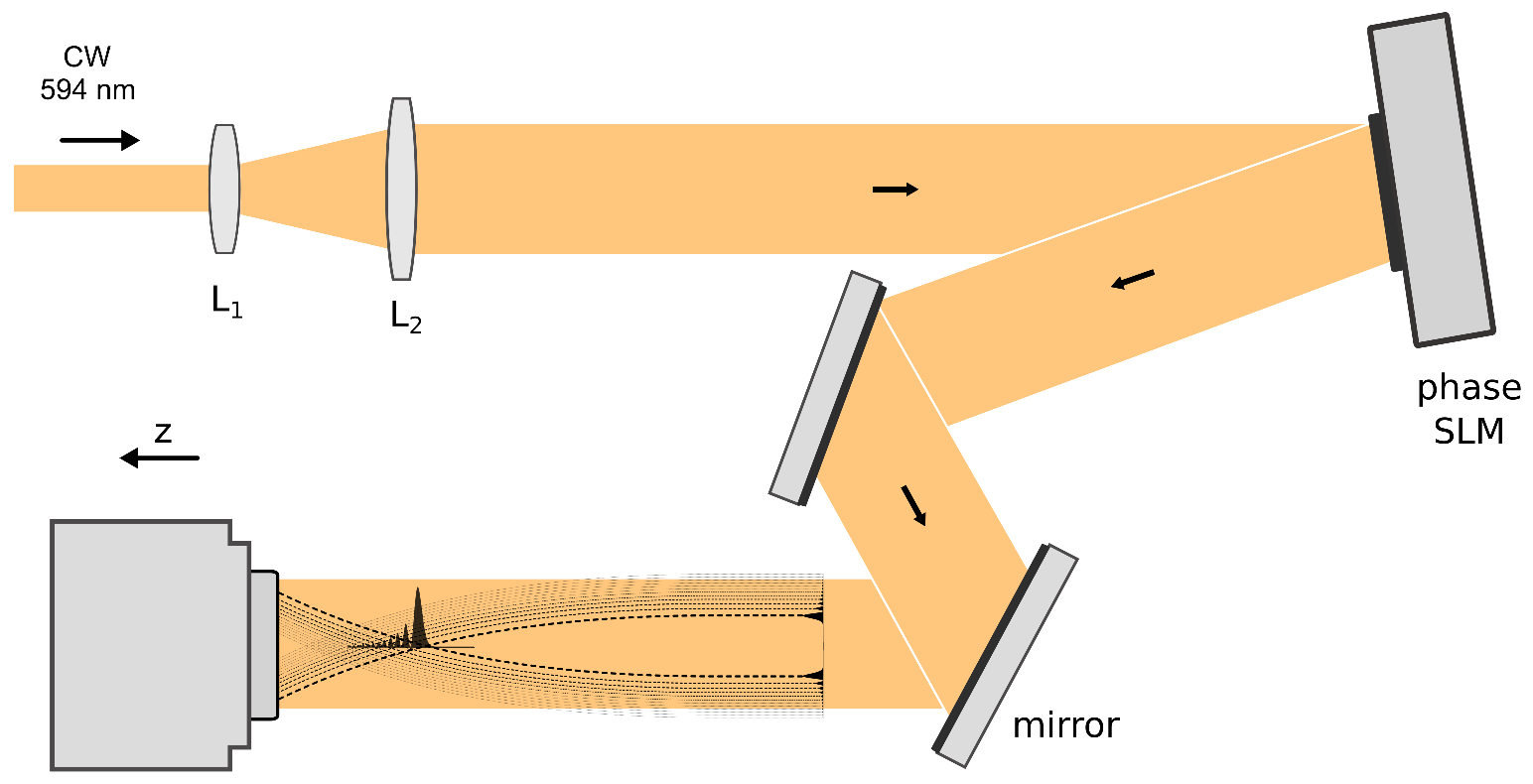}
\caption{Experimental setup for the generation of Tornado waves. Lenses $L_1, L_2$ comprise a 2x beam expander. SLM: 1920x1080 pixels, $8 ~\rm{\mu m}$ pixel size, CCD: 1280x1024 pixels, 8bit.}
\label{fig:setup}
\end{figure}
\begin{table}[b]
\centering
\caption{{Design parameters for the generation of ToWs. ($f$ refers to predicted values ) 
}}
\label{tab:MaskParameters}
\addtolength{\tabcolsep}{5pt} 
\begin{tabular}{cccccc}
\hline
\multirow{2}{*}{} & $r_o$  & $C$    &\multirow{2}{*}{$l$ }  & $f$   & $\lambda$ \\
                     & $(\mu\text{m})$ &      $({10^{ - 5}}/{\sqrt {\mu{\text{m}}} })$         &              & (mm)             & (nm)    \\ \hline
$\varPhi_A$          & 800            & 8.9            & +1,+2            & 424             & 594       \\
$\varPhi_B$          & 1000           & 10.0            &-1,-2            & 422             & 594       \\ \hline
\end{tabular}
\addtolength{\tabcolsep}{-5pt} 
\end{table}
\begin{figure}[t]
\centering
\includegraphics[width=0.45\textwidth]{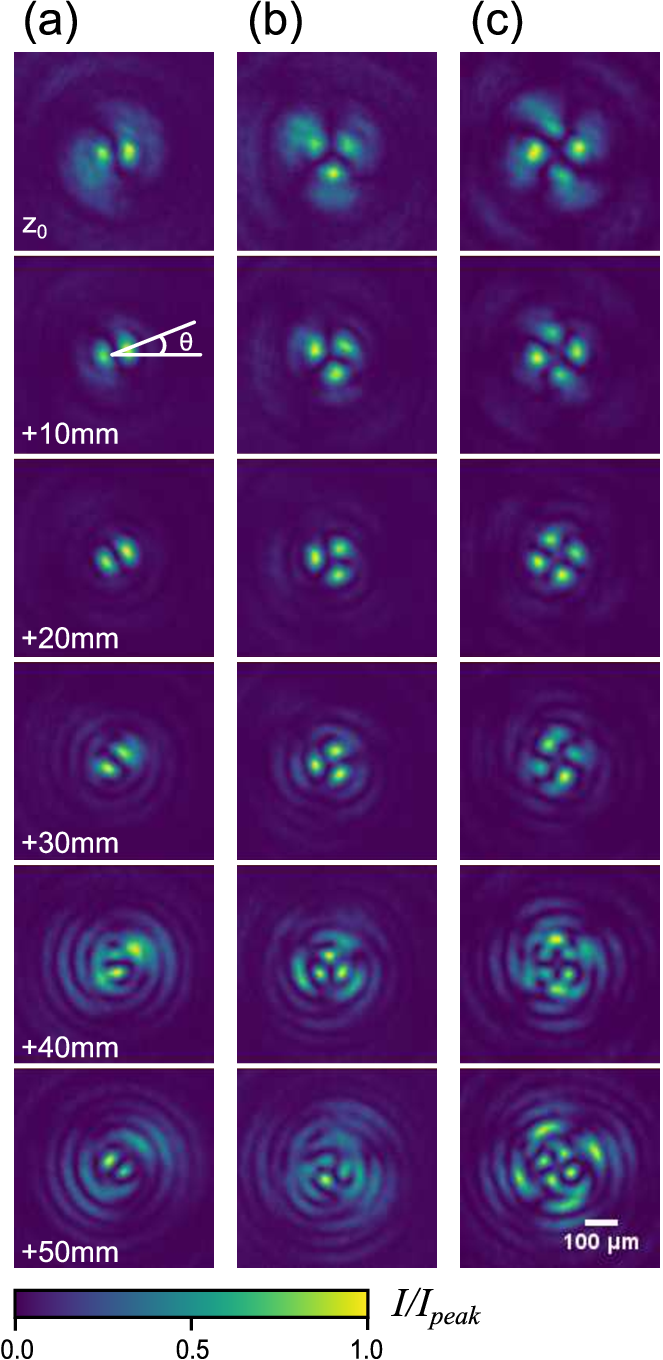}
\caption{Normalized transverse $I(x,y)$ intensity profiles (experimental results), recorded at different distances along propagation, of Tornado waves with 2, 3, and 4 lobes respectively (a) $l_A=1$, $l_B=-1$, (b) $l_A=1$, $l_B=-2$, (c) $l_A=2$, $l_B=-2$.}
\label{fig:stack}
\end{figure}
\begin{figure*}[]
\centering
\includegraphics[width=0.95\textwidth]{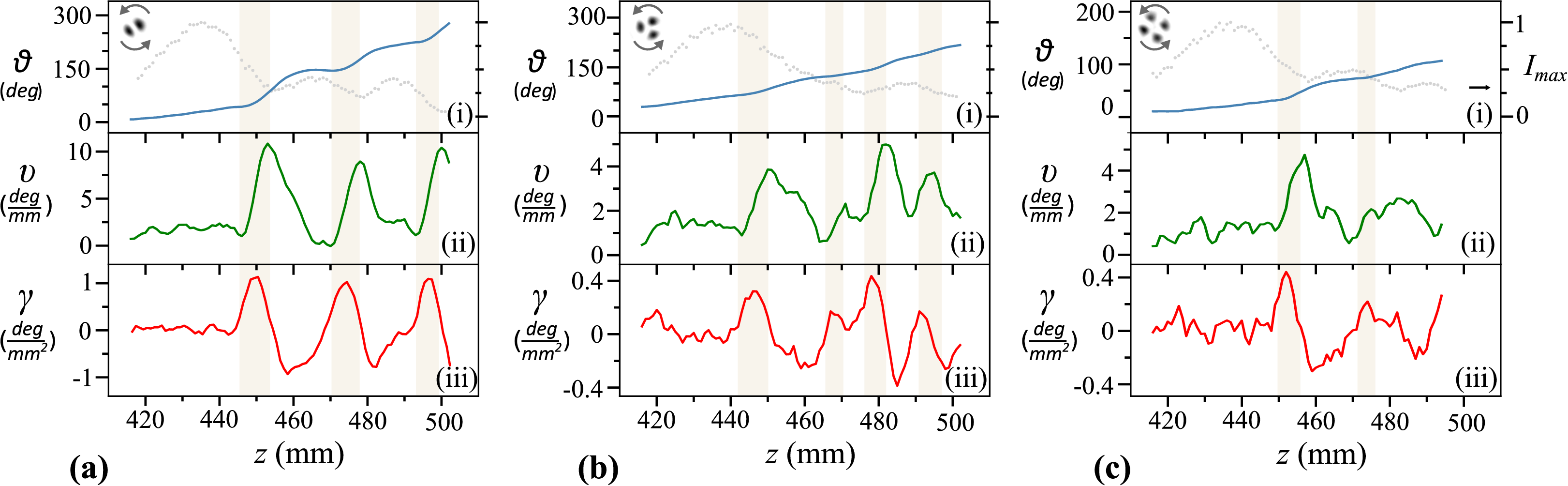}
\caption{Evolution of the lobes of Tornado waves along propagation (a) two lobes with $l_A=1,l_B=-1$, (b) three lobes with $l_A=1,l_B=-2$, and (c) four lobes with $l_A=2,l_B=-2$. Sub-figures: (i) lobe angle $\theta$ (solid line) and normalized maximum peak intensity (dotted line), (ii) angular velocity $v$, and (iii) angular acceleration $\gamma$.}
\label{fig:l1m1}
\end{figure*}
We applied this multiplexing approach to generate superimposing accelerating waves $u_A,~u_B$ carrying OAM. Furthermore, for each of the superimposed fields a phase mask was designed following an approach similar to \cite{Chremmos2011,Liu2016}. The phase in each phase mask can be described as $\varPhi\left(r,\theta\right)= \varphi\left(r\right)+\psi \left(\theta\right)$ where $\varphi \left( r \right) = -k C{\left( {r - {r_o}} \right)^{3/2}},~\forall r \ge {r_o} \wedge \varphi \left( r \right) = 0,~\forall r < {r_o}$ is  a radially chirped phase distribution, $\psi \left( \theta \right) = l\theta$, is a linear helical phase and $r$ is the radial coordinate, $r_o$ is a reference radius, $C$ is a constant, $\theta$ is the azimuthal angle, and  $l$ is the topological charge.

The design parameters we used for the generation of the OAM carrying, superimposing accelerating waves are shown in Table~\ref{tab:MaskParameters}. In our design we have tuned \cite{brimis2020tornado} the foci of $u_A, u_B$ to overlap. The position of the abrupt autofocus of a ToW can be estimated with good accuracy using the analytical solution of the one-dimensional Airy beam \cite{brimis2020tornado, Mansour2018,Papazoglou2011}. In our case, using the parabolic trajectory \cite{Cottrell2009} of the caustic resulting from phase distribution $\varPhi$, we estimate that the position of the abrupt autofocus is at $f = \frac{4}{3 C}\sqrt {r_o}$. The individual phase masks $\varPhi_A,~\varPhi_B$ were then spatially multiplexed to a single phase mask that was used in the SLM as shown in Fig. \ref{fig:PhaseMask}. 

In order to investigate the effectiveness of our approach for generating ToWs we performed numerical simulations based on  Eq.(\ref{eq:paraxial}). In our simulations we considered a monochromatic  ($\lambda=594~$nm) CW linearly polarized Gaussian beam along the $\textbf{y}$ direction that illuminates the mask which is shown in Fig. \ref{fig:PhaseMask}. The intensity profile along propagation direction is presented in Fig. \ref{fig:I_z}(a). The abruptly autofocusing characteristics are clearly reproduced. The diagonal stripes appearing in this $x-z$  cross section are due to the presence of the diffracted orders. A clear demonstration of the effectiveness of the spatial multiplexing approach is demonstrated in  Fig. \ref{fig:I_z}(b) where the transverse $I(x,y)$ intensity profile at $z=10,~20,~30~$cm is depicted. The Gaussian intensity profile of the propagating beam is gradually transformed to a vortex shape, a distinctive characteristic of ToW's \cite{brimis2020tornado}. Furthermore, besides the zero order, two replicas representing the $\pm 1$ diffraction orders, are clearly visible. Although initially overlapping at $z=10~$cm, due to their angular separation of $\sim2.13^o$ they are spatially separated at $z=30~$cm.

For the generation of Tornado waves we used a simple experimental setup as shown in Fig. \ref{fig:setup}. A CW laser beam of Gaussian profile and wavelength $\lambda=594~\rm{nm}$, {is expanded by a factor 2x using lenses $L_1, L_2$ to a Gaussian beam with $\mathrm{FWHM}=10.9~\rm{mm}$} and then illuminates a reflecting phase SLM. The modulated beam is then directed, using reflecting mirrors, towards a digital camera. In order to record the modulated beam intensity profile $I(x,y)$ along propagation, the camera is translated along the $z$ axis. Only the zero order is imaged since all higher diffraction orders are filtered out due to their angular and spatial separation from the zero order. 

{\it Results and Discussion.}---Using the experimental setup depicted in Fig. \ref{fig:setup} and applying the multiplexing technique we generated ToWs by superimposing accelerating waves carrying OAM with the design parameters presented in Table \ref{tab:MaskParameters}. 
In our experiments we used 3 combinations of topological charge $(l_{A,B}=\pm 1,~l_{A,B}=+1,-2,~l_{A,B}=+2,-2)$ resulting respectively into 2, 3 and 4 twisting intensity lobes. 

The evolution of the transverse intensity $I(x,y)$ distribution as a ToW propagates along $z$ is shown in  Fig. \ref{fig:stack}. Starting from a reference point $z_0=414~\rm{mm}$ along the propagation we can observe that the intensity lobes rotate at a varying rate and a decreasing radius around $z$ axis (located at the center of each image). This is a typical behavior of ToWs, where light twists and accelerates both in the angular and in the radial dimension \cite{brimis2020tornado}. The $I(x,y)$ intensity pattern becomes more complex as the number of lobes $N$ is increased  \cite{brimis2020tornado}. In all cases the high intensity lobes rotate at a varying rate and decreasing in radius around $z$ axis, although, as we observe by comparing Fig. \ref{fig:stack} (a) to (b) and (c), this becomes less profound as the number of lobes is increasing. 

In order to evaluate the accelerating characteristics of such ToWs we retrieved and analyzed the $I(x,y)$ cross sectional images {with a $\Delta z= 1~\rm{mm}$ sampling rate along z axis}. {We were able to monitor the trajectory of each twisting intensity lobe by monitoring it's radial and angular position. In more detail, first we identified the position of the maximum intensity for each lobe. From these points we accurately estimated the position of the rotation axis in our images. Using this as a point of reference along propagation we then tracked the radial and angular position of each lobe}. In Fig. \ref{fig:l1m1} we quantitatively depict the evolution of the angular orientation as a function of the propagation distance $z$ for the case of ToWs with 2, 3, and 4 high intensity lobes. In Figs. \ref{fig:l1m1} (a)-(c);(i) the angular position $\theta$ is depicted along with the normalized peak intensity. In all cases the angle $\theta$ increases at a varying rate as indicated from its oscillatory behavior. From these measured values we can estimate the angular velocity $v \equiv \dot \theta(z)$ and the angular acceleration $\gamma \equiv \ddot \theta(z)$ that are depicted in Figs. \ref{fig:l1m1}(a)-(c);(ii), (a)-(c);(iii) respectively. The vertical shaded areas highlight the zones where we observe angular acceleration. In all cases we observe a typical ToW behaviour, where the angular velocity $v$ varies in a quasi-periodic fashion between 0 and a peak value that ranges from $\sim  10~\rm{deg/mm}$ for the case of two lobes to $\sim  4~\rm{deg/mm}$ for the case of 4 lobes. As we can observe from Figs. \ref{fig:l1m1}(a)-(c);(iii) these areas are related to angular acceleration, where $\gamma$ takes values between $\sim  1.2~\rm{deg/mm}^2$ for the case of two lobes down to $\sim  0.4~\rm{deg/mm}$ for the case of 4 lobes. The estimated angular acceleration values are in good agreement with the values theoretically predicted \cite{brimis2020tornado} for ToWs in this range of autofocusing values. Interestingly, the peak values of angular acceleration are observed after the abrupt focus position, at areas where peak intensity of the twisting lobes drops, $\partial I/ \partial z <0$. Likewise, as $z$ is increased and the twisting lobes regain their intensity, angular deceleration takes place $(\gamma<0)$. For example, the transition from angular acceleration to deceleration is observed in \ref{fig:l1m1}(a);(iii) from $z \sim 450~ \rm{mm}$ to $z \sim 460 ~\rm{mm}$. This behaviour is related to the energy exchange between the ToW funnel and the reservoir regions \cite{brimis2020tornado}.  
\begin{figure}[b]
\centering
\includegraphics[width=0.7\textwidth]{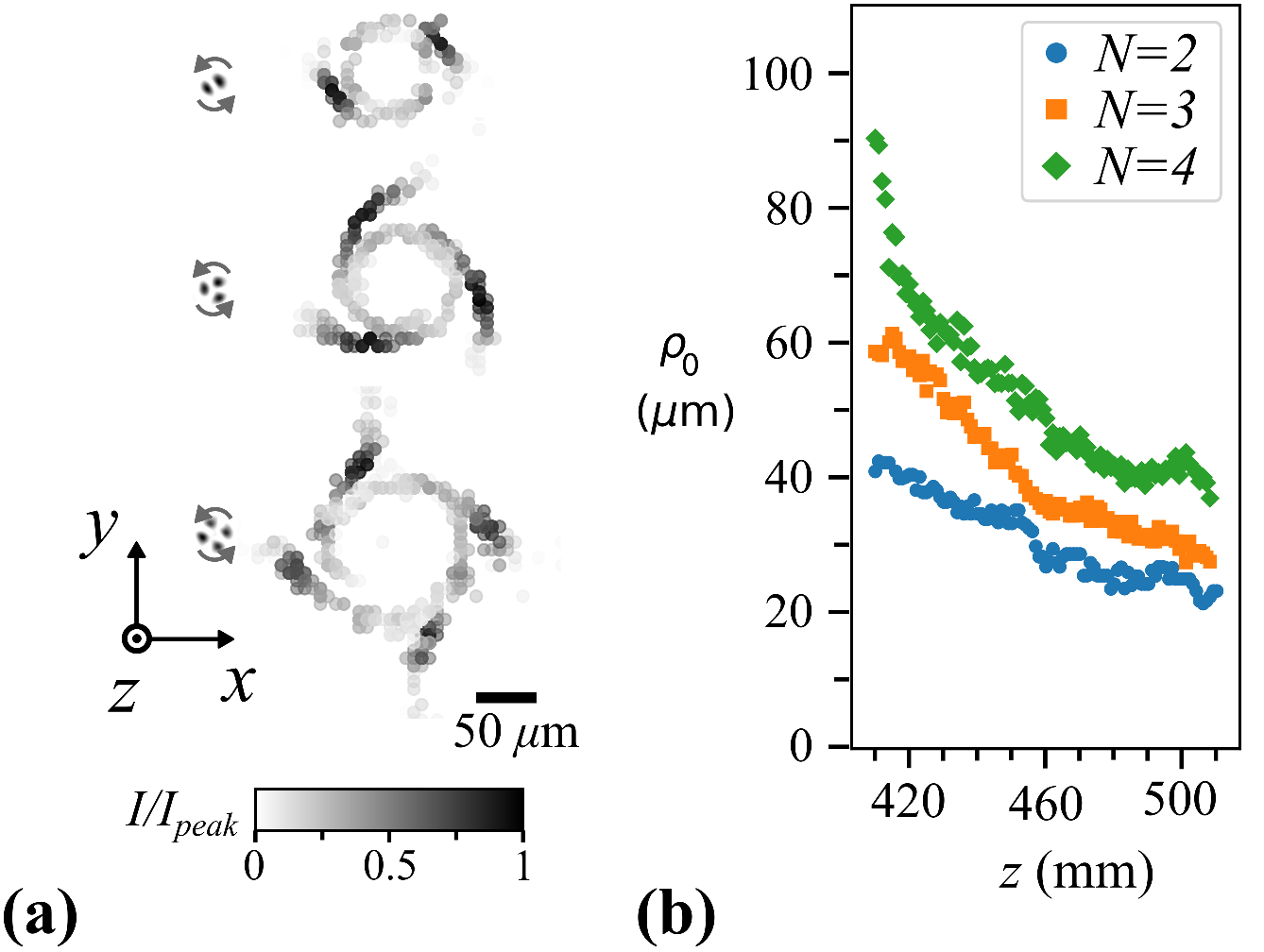}
\caption{{Trajectory of the high intensity lobes of Tornado waves for $N=2,~3,~4$. (a) Projection of the spiral trajectory along the propagation direction, on an $(x,y)$ plane. (The color of each point represents it's intensity normalized over the peak value). (b) Radial position $\rho_0$ of the high intensity lobes as a function of the propagation distance. (average value)}}
\label{fig:Rho_z}
\end{figure}
{The trajectory of the experimentally generated ToWs is visualized in Fig. \ref{fig:Rho_z}(a). By projecting the position of the high intensity lobes along the propagation direction, on a $(x,y)$ plane, we can clearly identify their tornado-like spiraling trajectory. The radial distance $\rho_0$ of the rotating high intensity lobes along propagation is shown in Fig. \ref{fig:Rho_z}(b) for ToWs with $N=2,~3,~4$ lobes respectively. Here we focus our attention in the ToWs funnel region \cite{brimis2020tornado}, and more specifically in the range $450 \rm{mm} < z < 500 \rm{mm}$ where, as we can see from Fig. \ref{fig:l1m1}, angular acceleration $\gamma$ is observed. In this region the rotating lobe's radial position $\rho_0$ shrinks by $30 \pm 3\%$. We note here that the non-linear parabolic trajectory of the radial distance is not profound in Fig. \ref{fig:Rho_z}(b), since we have isolated only a small part of it, near the abrupt autofocus \cite{Papazoglou2011, brimis2020tornado}. }

\begin{figure}[b]
\centering
\includegraphics[width=0.7\textwidth]{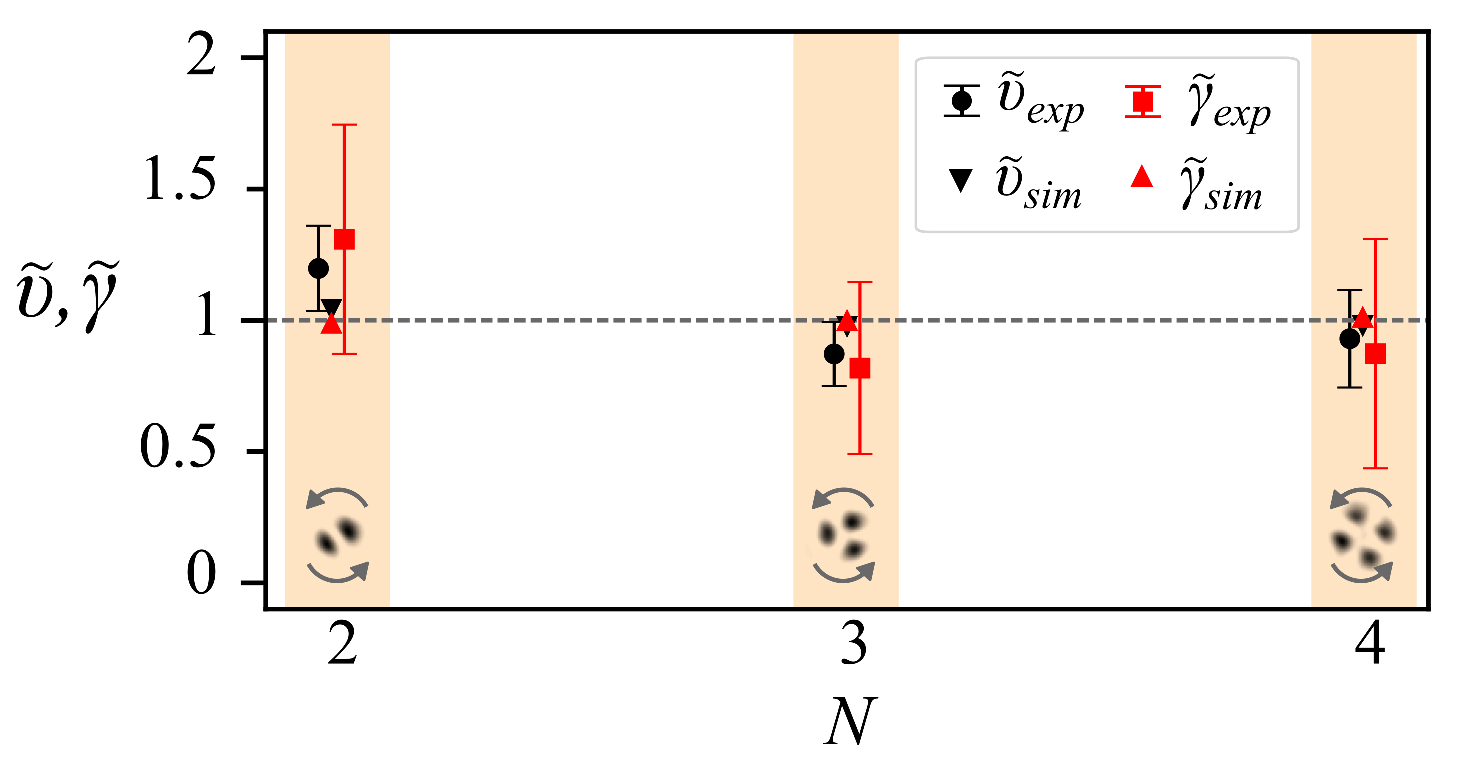}
\caption{Dependence of the angular velocity and acceleration as a function of the number of lobes  high intensity lobes. Comparison between normalized ${\tilde \upsilon},~  {\tilde \gamma} $ experimental measurements and simulations. The dashed line is the theoretical prediction.}
\label{fig:NScaling}
\end{figure}
Furthermore, both the peak intensity and the peak values for $v_{max}, \gamma_{max}$ decrease as the number of lobes $N$ is increased. The drop in the peak intensity is a direct outcome of distributing the beam's energy in the ToW funnel region \cite{brimis2020tornado} as the number of lobes is increasing. In order to test the validity of the $1/N$ dependence as described in Eqs.(\ref{eq:scaleangvel},\ref{eq:scaleAcc}) we have also performed numerical simulations of the propagation of ToWs that are generated by interfering ring-Airy beams with parameters described in \ref{tab:MaskParameters}. According to Eqs. (\ref{eq:scaleangvel}), (\ref{eq:scaleAcc}) the values of angular velocity $\upsilon$ and acceleration $\gamma$  are respectively proportional to $\partial \Delta \phi/\partial z$ and $\partial^2 \Delta \phi/\partial z^2$. These values are difficult to be estimated from our experiments, thus we have used a normalization scheme to study the dependence of $\upsilon, \gamma$ on the number of lobes $N$. In more detail, our experiments and simulations were performed in sets. In each set, we have changed the number of lobes by changing only the topological charges, keeping all other parameters the same (see Table \ref{tab:MaskParameters}). Firstly, for each case $i$ in a set we estimate $N_i \upsilon_i, N_i \gamma_i$, where $\upsilon_i, \gamma_i$ refer to the peak values as the beams propagate along the $z$ axis. Normalized values of angular velocity  ${\tilde \upsilon_i}$ and acceleration ${\tilde \gamma_i}$ are then retrieved by dividing these estimates with their average value  $\left\langle {N_i \upsilon_i} \right\rangle, \left\langle {N_i \gamma_i} \right\rangle$ for each set of measurements, i.e. ${\tilde \upsilon_i} \equiv N_i \upsilon_i/ \left\langle {N_i \upsilon_i} \right\rangle, {\tilde \gamma_i} \equiv {N_i \gamma_i}/ \left\langle {N_i \gamma_i} \right\rangle$. According to Eqs.(\ref{eq:scaleangvel},\ref{eq:scaleAcc}) ideally ${\tilde \upsilon_i}={\tilde \gamma_i}=1$. The dependence of ${\tilde \upsilon},  {\tilde \gamma}$ on the number of lobes $N$ for all our experimental measurements and simulations is shown in Fig. \ref{fig:NScaling}. The error bars in the experimental measurements are mainly a result of the differentiation of the discrete data. Simulations are in excellent agreement with the prediction of Eq.(\ref{eq:scaleangvel},\ref{eq:scaleAcc}). Likewise, the experimental measurements are, within the experimental error, in {fair} agreement with the scaling law. {We believe that the quality of the experimentally generated ToWs is the source of the observed deviations. In combination with the results presented in Fig. \ref{fig:l1m1}, we observe that as the number of lobes is increased, the experimentally generated ToW propagation dynamics deteriorate. Imperfections in the ToW generation process can lead to such a behaviour.} For example, increasing the topological charge will increase the azimuthal gradient of the phase distribution and could lead to aliasing due the discreteness of the SLM and the multiplexing method used for the ToW generation. In the case of the simulations this is not observed since analytic expressions are used to describe the ToWs as interfering OAM carrying ring-Airy beam.   

{\it Conclusions.}---%
In conclusion we have experimentally studied the propagation dynamics of light that spirals like a tornado. By spatial multiplexing a phase SLM, we have generated two superimposing abruptly auto-focusing waves, carrying OAM of opposite handedness. Likewise, we studied the dependence of the angular velocity and acceleration on the beam parameters, including the dependence on the number of high intensity lobes. Furthermore we have proposed a two-color scheme that makes it possible to generate dynamically twisting light that can rotate at $\rm{THz}$ frequencies. Our results may pave the way for various novel applications that rely on the spiralling behavior of ToWs, like direct laser writing, optical tweezers and nonlinear optics.

\begin{acknowledgments}
The authors would like to thank G. Zacharakis and M. Mylonakis for their assistance in the implementation of the experimental setup.
\end{acknowledgments}

\bibliographystyle{apsrev4-1}
\bibliography{GToWs.bib}

\providecommand{\noopsort}[1]{}\providecommand{\singleletter}[1]{#1}%
\begin{thebibliography}{32}%
\makeatletter
\providecommand \@ifxundefined [1]{%
 \@ifx{#1\undefined}
}%
\providecommand \@ifnum [1]{%
 \ifnum #1\expandafter \@firstoftwo
 \else \expandafter \@secondoftwo
 \fi
}%
\providecommand \@ifx [1]{%
 \ifx #1\expandafter \@firstoftwo
 \else \expandafter \@secondoftwo
 \fi
}%
\providecommand \natexlab [1]{#1}%
\providecommand \enquote  [1]{``#1''}%
\providecommand \bibnamefont  [1]{#1}%
\providecommand \bibfnamefont [1]{#1}%
\providecommand \citenamefont [1]{#1}%
\providecommand \href@noop [0]{\@secondoftwo}%
\providecommand \href [0]{\begingroup \@sanitize@url \@href}%
\providecommand \@href[1]{\@@startlink{#1}\@@href}%
\providecommand \@@href[1]{\endgroup#1\@@endlink}%
\providecommand \@sanitize@url [0]{\catcode `\\12\catcode `\$12\catcode
  `\&12\catcode `\#12\catcode `\^12\catcode `\_12\catcode `\%12\relax}%
\providecommand \@@startlink[1]{}%
\providecommand \@@endlink[0]{}%
\providecommand \url  [0]{\begingroup\@sanitize@url \@url }%
\providecommand \@url [1]{\endgroup\@href {#1}{\urlprefix }}%
\providecommand \urlprefix  [0]{URL }%
\providecommand \Eprint [0]{\href }%
\providecommand \doibase [0]{https://doi.org/}%
\providecommand \selectlanguage [0]{\@gobble}%
\providecommand \bibinfo  [0]{\@secondoftwo}%
\providecommand \bibfield  [0]{\@secondoftwo}%
\providecommand \translation [1]{[#1]}%
\providecommand \BibitemOpen [0]{}%
\providecommand \bibitemStop [0]{}%
\providecommand \bibitemNoStop [0]{.\EOS\space}%
\providecommand \EOS [0]{\spacefactor3000\relax}%
\providecommand \BibitemShut  [1]{\csname bibitem#1\endcsname}%
\let\auto@bib@innerbib\@empty
\bibitem [{\citenamefont {Rubinsztein-Dunlop}\ \emph
  {et~al.}(2017)\citenamefont {Rubinsztein-Dunlop}, \citenamefont {Forbes},
  \citenamefont {Berry}, \citenamefont {Dennis}, \citenamefont {Andrews},\ and\
  \citenamefont {et~al.}}]{Rubinsztein-Dunlop2017}%
  \BibitemOpen
  \bibfield  {author} {\bibinfo {author} {\bibfnamefont {H.}~\bibnamefont
  {Rubinsztein-Dunlop}}, \bibinfo {author} {\bibfnamefont {A.}~\bibnamefont
  {Forbes}}, \bibinfo {author} {\bibfnamefont {M.~V.}\ \bibnamefont {Berry}},
  \bibinfo {author} {\bibfnamefont {M.~R.}\ \bibnamefont {Dennis}}, \bibinfo
  {author} {\bibfnamefont {D.~L.}\ \bibnamefont {Andrews}},\ and\ \bibinfo
  {author} {\bibnamefont {et~al.}},\ }\bibfield  {title} {\bibinfo {title}
  {{Roadmap on structured light}},\ }\href
  {https://doi.org/10.1088/2040-8978/19/1/013001} {\bibfield  {journal}
  {\bibinfo  {journal} {Journal of Optics}\ }\textbf {\bibinfo {volume} {19}},\
  \bibinfo {pages} {013001} (\bibinfo {year} {2017})}\BibitemShut {NoStop}%
\bibitem [{\citenamefont {Papazoglou}\ \emph {et~al.}(2011)\citenamefont
  {Papazoglou}, \citenamefont {Efremidis}, \citenamefont {Christodoulides},\
  and\ \citenamefont {Tzortzakis}}]{Papazoglou2011}%
  \BibitemOpen
  \bibfield  {author} {\bibinfo {author} {\bibfnamefont {D.~G.}\ \bibnamefont
  {Papazoglou}}, \bibinfo {author} {\bibfnamefont {N.~K.}\ \bibnamefont
  {Efremidis}}, \bibinfo {author} {\bibfnamefont {D.~N.}\ \bibnamefont
  {Christodoulides}},\ and\ \bibinfo {author} {\bibfnamefont {S.}~\bibnamefont
  {Tzortzakis}},\ }\bibfield  {title} {\bibinfo {title} {{Observation of
  abruptly autofocusing waves}},\ }\href
  {http://ol.osa.org/abstract.cfm?URI=ol-36-10-1842} {\bibfield  {journal}
  {\bibinfo  {journal} {Optics Letters}\ }\textbf {\bibinfo {volume} {36}},\
  \bibinfo {pages} {1842} (\bibinfo {year} {2011})}\BibitemShut {NoStop}%
\bibitem [{\citenamefont {Panagiotopoulos}\ \emph {et~al.}(2013)\citenamefont
  {Panagiotopoulos}, \citenamefont {Papazoglou}, \citenamefont {Couairon},\
  and\ \citenamefont {Tzortzakis}}]{Panagiotopoulos2013}%
  \BibitemOpen
  \bibfield  {author} {\bibinfo {author} {\bibfnamefont {P.}~\bibnamefont
  {Panagiotopoulos}}, \bibinfo {author} {\bibfnamefont {D.}~\bibnamefont
  {Papazoglou}}, \bibinfo {author} {\bibfnamefont {A.}~\bibnamefont
  {Couairon}},\ and\ \bibinfo {author} {\bibfnamefont {S.}~\bibnamefont
  {Tzortzakis}},\ }\bibfield  {title} {\bibinfo {title} {{Sharply autofocused
  ring-Airy beams transforming into non-linear intense light bullets}},\ }\href
  {https://doi.org/10.1038/ncomms3622} {\bibfield  {journal} {\bibinfo
  {journal} {Nature Communications}\ }\textbf {\bibinfo {volume} {4}},\
  \bibinfo {pages} {2622} (\bibinfo {year} {2013})}\BibitemShut {NoStop}%
\bibitem [{\citenamefont {Froehly}\ \emph {et~al.}(2011)\citenamefont
  {Froehly}, \citenamefont {Courvoisier}, \citenamefont {Mathis}, \citenamefont
  {Jacquot}, \citenamefont {Furfaro}, \citenamefont {Giust}, \citenamefont
  {Lacourt},\ and\ \citenamefont {Dudley}}]{Froehly2011}%
  \BibitemOpen
  \bibfield  {author} {\bibinfo {author} {\bibfnamefont {L.}~\bibnamefont
  {Froehly}}, \bibinfo {author} {\bibfnamefont {F.}~\bibnamefont
  {Courvoisier}}, \bibinfo {author} {\bibfnamefont {A.}~\bibnamefont {Mathis}},
  \bibinfo {author} {\bibfnamefont {M.}~\bibnamefont {Jacquot}}, \bibinfo
  {author} {\bibfnamefont {L.}~\bibnamefont {Furfaro}}, \bibinfo {author}
  {\bibfnamefont {R.}~\bibnamefont {Giust}}, \bibinfo {author} {\bibfnamefont
  {P.~A.}\ \bibnamefont {Lacourt}},\ and\ \bibinfo {author} {\bibfnamefont
  {J.~M.}\ \bibnamefont {Dudley}},\ }\bibfield  {title} {\bibinfo {title}
  {{Arbitrary accelerating micron-scale caustic beams in two and three
  dimensions}},\ }\href
  {http://www.opticsexpress.org/abstract.cfm?URI=oe-19-17-16455} {\bibfield
  {journal} {\bibinfo  {journal} {Optics Express}\ }\textbf {\bibinfo {volume}
  {19}},\ \bibinfo {pages} {16455} (\bibinfo {year} {2011})}\BibitemShut
  {NoStop}%
\bibitem [{\citenamefont {Koulouklidis}\ \emph {et~al.}(2017)\citenamefont
  {Koulouklidis}, \citenamefont {Papazoglou}, \citenamefont {Fedorov},\ and\
  \citenamefont {Tzortzakis}}]{Koulouklidis2017}%
  \BibitemOpen
  \bibfield  {author} {\bibinfo {author} {\bibfnamefont {A.~D.}\ \bibnamefont
  {Koulouklidis}}, \bibinfo {author} {\bibfnamefont {D.~G.}\ \bibnamefont
  {Papazoglou}}, \bibinfo {author} {\bibfnamefont {V.~Y.}\ \bibnamefont
  {Fedorov}},\ and\ \bibinfo {author} {\bibfnamefont {S.}~\bibnamefont
  {Tzortzakis}},\ }\bibfield  {title} {\bibinfo {title} {{Phase Memory
  Preserving Harmonics from Abruptly Autofocusing Beams}},\ }\href
  {https://doi.org/10.1103/PhysRevLett.119.223901} {\bibfield  {journal}
  {\bibinfo  {journal} {Physical Review Letters}\ }\textbf {\bibinfo {volume}
  {119}},\ \bibinfo {pages} {223901} (\bibinfo {year} {2017})},\ \Eprint
  {https://arxiv.org/abs/1902.07991} {1902.07991} \BibitemShut {NoStop}%
\bibitem [{\citenamefont {Liu}\ \emph {et~al.}(2016)\citenamefont {Liu},
  \citenamefont {Koulouklidis}, \citenamefont {Papazoglou}, \citenamefont
  {Tzortzakis},\ and\ \citenamefont {Zhang}}]{Liu2016}%
  \BibitemOpen
  \bibfield  {author} {\bibinfo {author} {\bibfnamefont {K.}~\bibnamefont
  {Liu}}, \bibinfo {author} {\bibfnamefont {A.~D.}\ \bibnamefont
  {Koulouklidis}}, \bibinfo {author} {\bibfnamefont {D.~G.}\ \bibnamefont
  {Papazoglou}}, \bibinfo {author} {\bibfnamefont {S.}~\bibnamefont
  {Tzortzakis}},\ and\ \bibinfo {author} {\bibfnamefont {X.-C.}\ \bibnamefont
  {Zhang}},\ }\bibfield  {title} {\bibinfo {title} {{Enhanced terahertz wave
  emission from air-plasma tailored by abruptly autofocusing laser beams}},\
  }\href {https://doi.org/10.1364/OPTICA.3.000605} {\bibfield  {journal}
  {\bibinfo  {journal} {Optica}\ }\textbf {\bibinfo {volume} {3}},\ \bibinfo
  {pages} {605} (\bibinfo {year} {2016})}\BibitemShut {NoStop}%
\bibitem [{\citenamefont {Couairon}\ and\ \citenamefont
  {Mysyrowicz}(2007)}]{Couairon2007b}%
  \BibitemOpen
  \bibfield  {author} {\bibinfo {author} {\bibfnamefont {A.}~\bibnamefont
  {Couairon}}\ and\ \bibinfo {author} {\bibfnamefont {A.}~\bibnamefont
  {Mysyrowicz}},\ }\bibfield  {title} {\bibinfo {title} {{Femtosecond
  filamentation in transparent media}},\ }\href
  {https://doi.org/10.1016/j.physrep.2006.12.005} {\bibfield  {journal}
  {\bibinfo  {journal} {Physics Reports}\ }\textbf {\bibinfo {volume} {441}},\
  \bibinfo {pages} {47} (\bibinfo {year} {2007})}\BibitemShut {NoStop}%
\bibitem [{\citenamefont {Siviloglou}\ \emph {et~al.}(2007)\citenamefont
  {Siviloglou}, \citenamefont {Broky}, \citenamefont {Dogariu},\ and\
  \citenamefont {Christodoulides}}]{Siviloglou2007}%
  \BibitemOpen
  \bibfield  {author} {\bibinfo {author} {\bibfnamefont {G.~A.}\ \bibnamefont
  {Siviloglou}}, \bibinfo {author} {\bibfnamefont {J.}~\bibnamefont {Broky}},
  \bibinfo {author} {\bibfnamefont {A.}~\bibnamefont {Dogariu}},\ and\ \bibinfo
  {author} {\bibfnamefont {D.~N.}\ \bibnamefont {Christodoulides}},\ }\bibfield
   {title} {\bibinfo {title} {{Observation of Accelerating Airy Beams}},\
  }\href {http://link.aps.org/abstract/PRL/v99/e213901
  http://dx.doi.org/10.1103/PhysRevLett.99.213901} {\bibfield  {journal}
  {\bibinfo  {journal} {Physical Review Letters}\ }\textbf {\bibinfo {volume}
  {99}},\ \bibinfo {pages} {213901} (\bibinfo {year} {2007})}\BibitemShut
  {NoStop}%
\bibitem [{\citenamefont {Efremidis}\ and\ \citenamefont
  {Christodoulides}(2010)}]{Efremidis2010a}%
  \BibitemOpen
  \bibfield  {author} {\bibinfo {author} {\bibfnamefont {N.~K.}\ \bibnamefont
  {Efremidis}}\ and\ \bibinfo {author} {\bibfnamefont {D.~N.}\ \bibnamefont
  {Christodoulides}},\ }\bibfield  {title} {\bibinfo {title} {{Abruptly
  autofocusing waves}},\ }\href {https://doi.org/10.1364/OL.35.004045}
  {\bibfield  {journal} {\bibinfo  {journal} {Optics Letters}\ }\textbf
  {\bibinfo {volume} {35}},\ \bibinfo {pages} {4045} (\bibinfo {year}
  {2010})}\BibitemShut {NoStop}%
\bibitem [{\citenamefont {Chremmos}\ \emph {et~al.}(2011)\citenamefont
  {Chremmos}, \citenamefont {Efremidis},\ and\ \citenamefont
  {Christodoulides}}]{Chremmos2011}%
  \BibitemOpen
  \bibfield  {author} {\bibinfo {author} {\bibfnamefont {I.}~\bibnamefont
  {Chremmos}}, \bibinfo {author} {\bibfnamefont {N.~K.}\ \bibnamefont
  {Efremidis}},\ and\ \bibinfo {author} {\bibfnamefont {D.~N.}\ \bibnamefont
  {Christodoulides}},\ }\bibfield  {title} {\bibinfo {title} {Pre-engineered
  abruptly autofocusing beams},\ }\href {https://doi.org/10.1364/OL.36.001890}
  {\bibfield  {journal} {\bibinfo  {journal} {Opt. Lett.}\ }\textbf {\bibinfo
  {volume} {36}},\ \bibinfo {pages} {1890} (\bibinfo {year}
  {2011})}\BibitemShut {NoStop}%
\bibitem [{\citenamefont {Durnin}\ \emph {et~al.}(1987)\citenamefont {Durnin},
  \citenamefont {Miceli},\ and\ \citenamefont {Eberly}}]{Durnin1987a}%
  \BibitemOpen
  \bibfield  {author} {\bibinfo {author} {\bibfnamefont {J.}~\bibnamefont
  {Durnin}}, \bibinfo {author} {\bibfnamefont {J.~J.}\ \bibnamefont {Miceli}},\
  and\ \bibinfo {author} {\bibfnamefont {J.~H.}\ \bibnamefont {Eberly}},\
  }\bibfield  {title} {\bibinfo {title} {{Diffraction-free beams}},\ }\href
  {http://link.aps.org/doi/10.1103/PhysRevLett.58.1499} {\bibfield  {journal}
  {\bibinfo  {journal} {Physical Review Letters}\ }\textbf {\bibinfo {volume}
  {58}},\ \bibinfo {pages} {1499} (\bibinfo {year} {1987})}\BibitemShut
  {NoStop}%
\bibitem [{\citenamefont {Carmon}\ \emph {et~al.}(2001)\citenamefont {Carmon},
  \citenamefont {Uzdin}, \citenamefont {Pigier}, \citenamefont {Musslimani},
  \citenamefont {Segev},\ and\ \citenamefont {Nepomnyashchy}}]{Carmon2001}%
  \BibitemOpen
  \bibfield  {author} {\bibinfo {author} {\bibfnamefont {T.}~\bibnamefont
  {Carmon}}, \bibinfo {author} {\bibfnamefont {R.}~\bibnamefont {Uzdin}},
  \bibinfo {author} {\bibfnamefont {C.}~\bibnamefont {Pigier}}, \bibinfo
  {author} {\bibfnamefont {Z.~H.}\ \bibnamefont {Musslimani}}, \bibinfo
  {author} {\bibfnamefont {M.}~\bibnamefont {Segev}},\ and\ \bibinfo {author}
  {\bibfnamefont {A.}~\bibnamefont {Nepomnyashchy}},\ }\bibfield  {title}
  {\bibinfo {title} {{Rotating Propeller Solitons}},\ }\href
  {https://doi.org/10.1103/PhysRevLett.87.143901} {\bibfield  {journal}
  {\bibinfo  {journal} {Physical Review Letters}\ }\textbf {\bibinfo {volume}
  {87}},\ \bibinfo {pages} {143901} (\bibinfo {year} {2001})}\BibitemShut
  {NoStop}%
\bibitem [{\citenamefont {Schulze}\ \emph {et~al.}(2015)\citenamefont
  {Schulze}, \citenamefont {Roux}, \citenamefont {Dudley}, \citenamefont {Rop},
  \citenamefont {Duparr{\'{e}}},\ and\ \citenamefont {Forbes}}]{Schulze2015}%
  \BibitemOpen
  \bibfield  {author} {\bibinfo {author} {\bibfnamefont {C.}~\bibnamefont
  {Schulze}}, \bibinfo {author} {\bibfnamefont {F.~S.}\ \bibnamefont {Roux}},
  \bibinfo {author} {\bibfnamefont {A.}~\bibnamefont {Dudley}}, \bibinfo
  {author} {\bibfnamefont {R.}~\bibnamefont {Rop}}, \bibinfo {author}
  {\bibfnamefont {M.}~\bibnamefont {Duparr{\'{e}}}},\ and\ \bibinfo {author}
  {\bibfnamefont {A.}~\bibnamefont {Forbes}},\ }\bibfield  {title} {\bibinfo
  {title} {{Accelerated rotation with orbital angular momentum modes}},\ }\href
  {https://doi.org/10.1103/PhysRevA.91.043821} {\bibfield  {journal} {\bibinfo
  {journal} {Physical Review A}\ }\textbf {\bibinfo {volume} {91}},\ \bibinfo
  {pages} {043821} (\bibinfo {year} {2015})}\BibitemShut {NoStop}%
\bibitem [{\citenamefont {Webster}\ \emph {et~al.}(2017)\citenamefont
  {Webster}, \citenamefont {Rosales-Guzm{\'{a}}n},\ and\ \citenamefont
  {Forbes}}]{Webster2017}%
  \BibitemOpen
  \bibfield  {author} {\bibinfo {author} {\bibfnamefont {J.}~\bibnamefont
  {Webster}}, \bibinfo {author} {\bibfnamefont {C.}~\bibnamefont
  {Rosales-Guzm{\'{a}}n}},\ and\ \bibinfo {author} {\bibfnamefont
  {A.}~\bibnamefont {Forbes}},\ }\bibfield  {title} {\bibinfo {title}
  {{Radially dependent angular acceleration of twisted light}},\ }\href
  {https://doi.org/10.1364/OL.42.000675} {\bibfield  {journal} {\bibinfo
  {journal} {Optics Letters}\ }\textbf {\bibinfo {volume} {42}},\ \bibinfo
  {pages} {675} (\bibinfo {year} {2017})}\BibitemShut {NoStop}%
\bibitem [{\citenamefont {Brimis}\ \emph {et~al.}(2020)\citenamefont {Brimis},
  \citenamefont {Makris},\ and\ \citenamefont
  {Papazoglou}}]{brimis2020tornado}%
  \BibitemOpen
  \bibfield  {author} {\bibinfo {author} {\bibfnamefont {A.}~\bibnamefont
  {Brimis}}, \bibinfo {author} {\bibfnamefont {K.~G.}\ \bibnamefont {Makris}},\
  and\ \bibinfo {author} {\bibfnamefont {D.~G.}\ \bibnamefont {Papazoglou}},\
  }\bibfield  {title} {\bibinfo {title} {Tornado waves},\ }\href@noop {}
  {\bibfield  {journal} {\bibinfo  {journal} {Optics Letters}\ }\textbf
  {\bibinfo {volume} {45}},\ \bibinfo {pages} {280} (\bibinfo {year}
  {2020})}\BibitemShut {NoStop}%
\bibitem [{\citenamefont {Mansour}\ \emph {et~al.}(2021)\citenamefont
  {Mansour}, \citenamefont {Brimis}, \citenamefont {Makris},\ and\
  \citenamefont {Papazoglou}}]{Mansour2021}%
  \BibitemOpen
  \bibfield  {author} {\bibinfo {author} {\bibfnamefont {D.}~\bibnamefont
  {Mansour}}, \bibinfo {author} {\bibfnamefont {A.}~\bibnamefont {Brimis}},
  \bibinfo {author} {\bibfnamefont {K.~G.}\ \bibnamefont {Makris}},\ and\
  \bibinfo {author} {\bibfnamefont {D.~G.}\ \bibnamefont {Papazoglou}},\
  }\bibfield  {title} {\bibinfo {title} {{Generation of Tornado Waves}},\ }in\
  \href {https://doi.org/10.1364/CLEO_QELS.2021.FTh1J.7} {\emph {\bibinfo
  {booktitle} {Conf. Lasers Electro-Optics}}}\ (\bibinfo  {publisher} {OSA},\
  \bibinfo {address} {Washington, D.C.},\ \bibinfo {year} {2021})\ p.\ \bibinfo
  {pages} {FTh1J.7}\BibitemShut {NoStop}%
\bibitem [{\citenamefont {Brabec}\ and\ \citenamefont
  {Krausz}(1997)}]{Brabec1997}%
  \BibitemOpen
  \bibfield  {author} {\bibinfo {author} {\bibfnamefont {T.}~\bibnamefont
  {Brabec}}\ and\ \bibinfo {author} {\bibfnamefont {F.}~\bibnamefont
  {Krausz}},\ }\bibfield  {title} {\bibinfo {title} {{Nonlinear Optical Pulse
  Propagation in the Single-Cycle Regime}},\ }\href
  {https://doi.org/10.1103/PhysRevLett.78.3282} {\bibfield  {journal} {\bibinfo
   {journal} {Phys. Rev. Lett.}\ }\textbf {\bibinfo {volume} {78}},\ \bibinfo
  {pages} {3282} (\bibinfo {year} {1997})}\BibitemShut {NoStop}%
\bibitem [{\citenamefont {Kotlyar}\ \emph {et~al.}(1997)\citenamefont
  {Kotlyar}, \citenamefont {Soǐfer},\ and\ \citenamefont
  {Khonina}}]{Kotlyar1997}%
  \BibitemOpen
  \bibfield  {author} {\bibinfo {author} {\bibfnamefont {V.~V.}\ \bibnamefont
  {Kotlyar}}, \bibinfo {author} {\bibfnamefont {V.~A.}\ \bibnamefont
  {Soǐfer}},\ and\ \bibinfo {author} {\bibfnamefont {S.~N.}\ \bibnamefont
  {Khonina}},\ }\bibfield  {title} {\bibinfo {title} {{Rotation of multimode
  Gauss-Laguerre light beams in free space}},\ }\href
  {https://doi.org/10.1134/1.1261648} {\bibfield  {journal} {\bibinfo
  {journal} {Technical Physics Letters}\ }\textbf {\bibinfo {volume} {23}},\
  \bibinfo {pages} {657} (\bibinfo {year} {1997})}\BibitemShut {NoStop}%
\bibitem [{\citenamefont {Kotlyar}\ \emph {et~al.}(2007)\citenamefont
  {Kotlyar}, \citenamefont {Khonina}, \citenamefont {Skidanov},\ and\
  \citenamefont {Soifer}}]{Kotlyar2007}%
  \BibitemOpen
  \bibfield  {author} {\bibinfo {author} {\bibfnamefont {V.~V.}\ \bibnamefont
  {Kotlyar}}, \bibinfo {author} {\bibfnamefont {S.~N.}\ \bibnamefont
  {Khonina}}, \bibinfo {author} {\bibfnamefont {R.~V.}\ \bibnamefont
  {Skidanov}},\ and\ \bibinfo {author} {\bibfnamefont {V.~A.}\ \bibnamefont
  {Soifer}},\ }\bibfield  {title} {\bibinfo {title} {{Rotation of laser beams
  with zero of the orbital angular momentum}},\ }\href
  {https://doi.org/10.1016/j.optcom.2007.01.059} {\bibfield  {journal}
  {\bibinfo  {journal} {Optics Communications}\ }\textbf {\bibinfo {volume}
  {274}},\ \bibinfo {pages} {8} (\bibinfo {year} {2007})}\BibitemShut {NoStop}%
\bibitem [{\citenamefont {Odoulov}\ \emph {et~al.}(2015)\citenamefont
  {Odoulov}, \citenamefont {Shumelyuk}, \citenamefont {Badorreck},
  \citenamefont {Nolte}, \citenamefont {Voit},\ and\ \citenamefont
  {Imlau}}]{Odoulov2015}%
  \BibitemOpen
  \bibfield  {author} {\bibinfo {author} {\bibfnamefont {S.}~\bibnamefont
  {Odoulov}}, \bibinfo {author} {\bibfnamefont {A.}~\bibnamefont {Shumelyuk}},
  \bibinfo {author} {\bibfnamefont {H.}~\bibnamefont {Badorreck}}, \bibinfo
  {author} {\bibfnamefont {S.}~\bibnamefont {Nolte}}, \bibinfo {author}
  {\bibfnamefont {K.-M.}\ \bibnamefont {Voit}},\ and\ \bibinfo {author}
  {\bibfnamefont {M.}~\bibnamefont {Imlau}},\ }\bibfield  {title} {\bibinfo
  {title} {{Interference and holography with femtosecond laser pulses of
  different colours}},\ }\href {https://doi.org/10.1038/ncomms6866} {\bibfield
  {journal} {\bibinfo  {journal} {Nature Communications}\ }\textbf {\bibinfo
  {volume} {6}},\ \bibinfo {pages} {5866} (\bibinfo {year} {2015})}\BibitemShut
  {NoStop}%
\bibitem [{\citenamefont {Rego}\ \emph {et~al.}(2019)\citenamefont {Rego},
  \citenamefont {Dorney}, \citenamefont {Brooks}, \citenamefont {Nguyen},
  \citenamefont {Liao}, \citenamefont {Rom{\'{a}}n}, \citenamefont {Couch},
  \citenamefont {Liu}, \citenamefont {Pisanty}, \citenamefont {Lewenstein},
  \citenamefont {Plaja}, \citenamefont {Kapteyn}, \citenamefont {Murnane},\
  and\ \citenamefont {Hern{\'{a}}ndez-Garc{\'{i}}a}}]{Rego2019}%
  \BibitemOpen
  \bibfield  {author} {\bibinfo {author} {\bibfnamefont {L.}~\bibnamefont
  {Rego}}, \bibinfo {author} {\bibfnamefont {K.~M.}\ \bibnamefont {Dorney}},
  \bibinfo {author} {\bibfnamefont {N.~J.}\ \bibnamefont {Brooks}}, \bibinfo
  {author} {\bibfnamefont {Q.~L.}\ \bibnamefont {Nguyen}}, \bibinfo {author}
  {\bibfnamefont {C.~T.}\ \bibnamefont {Liao}}, \bibinfo {author}
  {\bibfnamefont {J.~S.}\ \bibnamefont {Rom{\'{a}}n}}, \bibinfo {author}
  {\bibfnamefont {D.~E.}\ \bibnamefont {Couch}}, \bibinfo {author}
  {\bibfnamefont {A.}~\bibnamefont {Liu}}, \bibinfo {author} {\bibfnamefont
  {E.}~\bibnamefont {Pisanty}}, \bibinfo {author} {\bibfnamefont
  {M.}~\bibnamefont {Lewenstein}}, \bibinfo {author} {\bibfnamefont
  {L.}~\bibnamefont {Plaja}}, \bibinfo {author} {\bibfnamefont {H.~C.}\
  \bibnamefont {Kapteyn}}, \bibinfo {author} {\bibfnamefont {M.~M.}\
  \bibnamefont {Murnane}},\ and\ \bibinfo {author} {\bibfnamefont
  {C.}~\bibnamefont {Hern{\'{a}}ndez-Garc{\'{i}}a}},\ }\bibfield  {title}
  {\bibinfo {title} {{Generation of extreme-ultraviolet beams with time-varying
  orbital angular momentum}},\ }\bibfield  {journal} {\bibinfo  {journal}
  {Science (80-. ).}\ }\textbf {\bibinfo {volume} {364}},\ \href
  {https://doi.org/10.1126/science.aaw9486} {10.1126/science.aaw9486} (\bibinfo
  {year} {2019})\BibitemShut {NoStop}%
\bibitem [{\citenamefont {Zhao}\ \emph {et~al.}(2020)\citenamefont {Zhao},
  \citenamefont {Song}, \citenamefont {Zhang}, \citenamefont {Pang},
  \citenamefont {Liu}, \citenamefont {Song}, \citenamefont {Almaiman},
  \citenamefont {Manukyan}, \citenamefont {Zhou}, \citenamefont {Lynn},
  \citenamefont {Boyd}, \citenamefont {Tur},\ and\ \citenamefont
  {Willner}}]{Zhao2020}%
  \BibitemOpen
  \bibfield  {author} {\bibinfo {author} {\bibfnamefont {Z.}~\bibnamefont
  {Zhao}}, \bibinfo {author} {\bibfnamefont {H.}~\bibnamefont {Song}}, \bibinfo
  {author} {\bibfnamefont {R.}~\bibnamefont {Zhang}}, \bibinfo {author}
  {\bibfnamefont {K.}~\bibnamefont {Pang}}, \bibinfo {author} {\bibfnamefont
  {C.}~\bibnamefont {Liu}}, \bibinfo {author} {\bibfnamefont {H.}~\bibnamefont
  {Song}}, \bibinfo {author} {\bibfnamefont {A.}~\bibnamefont {Almaiman}},
  \bibinfo {author} {\bibfnamefont {K.}~\bibnamefont {Manukyan}}, \bibinfo
  {author} {\bibfnamefont {H.}~\bibnamefont {Zhou}}, \bibinfo {author}
  {\bibfnamefont {B.}~\bibnamefont {Lynn}}, \bibinfo {author} {\bibfnamefont
  {R.~W.}\ \bibnamefont {Boyd}}, \bibinfo {author} {\bibfnamefont
  {M.}~\bibnamefont {Tur}},\ and\ \bibinfo {author} {\bibfnamefont {A.~E.}\
  \bibnamefont {Willner}},\ }\bibfield  {title} {\bibinfo {title} {{Dynamic
  spatiotemporal beams that combine two independent and controllable
  orbital-angular-momenta using multiple optical-frequency-comb lines}},\
  }\href {https://doi.org/10.1038/s41467-020-17805-1} {\bibfield  {journal}
  {\bibinfo  {journal} {Nat. Commun.}\ }\textbf {\bibinfo {volume} {11}},\
  \bibinfo {pages} {4099} (\bibinfo {year} {2020})},\ \Eprint
  {https://arxiv.org/abs/1904.13150} {arXiv:1904.13150} \BibitemShut {NoStop}%
\bibitem [{\citenamefont {Béjot}\ and\ \citenamefont
  {Kibler}(2021)}]{Bejot2021}%
  \BibitemOpen
  \bibfield  {author} {\bibinfo {author} {\bibfnamefont {P.}~\bibnamefont
  {Béjot}}\ and\ \bibinfo {author} {\bibfnamefont {B.}~\bibnamefont
  {Kibler}},\ }\bibfield  {title} {\bibinfo {title} {Spatiotemporal helicon
  wavepackets},\ }\href {https://doi.org/10.1021/acsphotonics.1c00522}
  {\bibfield  {journal} {\bibinfo  {journal} {ACS Photonics}\ }\textbf
  {\bibinfo {volume} {8}},\ \bibinfo {pages} {2345} (\bibinfo {year} {2021})},\
  \Eprint {https://arxiv.org/abs/https://doi.org/10.1021/acsphotonics.1c00522}
  {https://doi.org/10.1021/acsphotonics.1c00522} \BibitemShut {NoStop}%
\bibitem [{\citenamefont {Wang}\ and\ \citenamefont {Liang}(2021)}]{Wang2021}%
  \BibitemOpen
  \bibfield  {author} {\bibinfo {author} {\bibfnamefont {J.}~\bibnamefont
  {Wang}}\ and\ \bibinfo {author} {\bibfnamefont {Y.}~\bibnamefont {Liang}},\
  }\bibfield  {title} {\bibinfo {title} {{Generation and Detection of
  Structured Light: A Review}},\ }\href
  {https://doi.org/10.3389/fphy.2021.688284} {\bibfield  {journal} {\bibinfo
  {journal} {Front. Phys.}\ }\textbf {\bibinfo {volume} {9}},\ \bibinfo {pages}
  {1} (\bibinfo {year} {2021})}\BibitemShut {NoStop}%
\bibitem [{\citenamefont {Goodman}(1996)}]{Goodman1996}%
  \BibitemOpen
  \bibfield  {author} {\bibinfo {author} {\bibfnamefont {J.~W.}\ \bibnamefont
  {Goodman}},\ }\href@noop {} {\emph {\bibinfo {title} {{Introduction to
  Fourier Optics}}}},\ \bibinfo {edition} {2nd}\ ed.\ (\bibinfo  {publisher}
  {McGraw-Hill},\ \bibinfo {address} {New York},\ \bibinfo {year}
  {1996})\BibitemShut {NoStop}%
\bibitem [{\citenamefont {Arriz{\'{o}}n}\ \emph {et~al.}(2007)\citenamefont
  {Arriz{\'{o}}n}, \citenamefont {Ruiz}, \citenamefont {Carrada},\ and\
  \citenamefont {Gonz{\'{a}}lez}}]{Arrizon2007}%
  \BibitemOpen
  \bibfield  {author} {\bibinfo {author} {\bibfnamefont {V.}~\bibnamefont
  {Arriz{\'{o}}n}}, \bibinfo {author} {\bibfnamefont {U.}~\bibnamefont {Ruiz}},
  \bibinfo {author} {\bibfnamefont {R.}~\bibnamefont {Carrada}},\ and\ \bibinfo
  {author} {\bibfnamefont {L.~A.}\ \bibnamefont {Gonz{\'{a}}lez}},\ }\bibfield
  {title} {\bibinfo {title} {{Pixelated phase computer holograms for the
  accurate encoding of scalar complex fields}},\ }\href
  {https://doi.org/10.1364/JOSAA.24.003500} {\bibfield  {journal} {\bibinfo
  {journal} {J. Opt. Soc. Am. A}\ }\textbf {\bibinfo {volume} {24}},\ \bibinfo
  {pages} {3500} (\bibinfo {year} {2007})}\BibitemShut {NoStop}%
\bibitem [{\citenamefont {Mendoza-Yero}\ \emph {et~al.}(2014)\citenamefont
  {Mendoza-Yero}, \citenamefont {M{\'{i}}nguez-Vega},\ and\ \citenamefont
  {Lancis}}]{Mendoza-Yero2014}%
  \BibitemOpen
  \bibfield  {author} {\bibinfo {author} {\bibfnamefont {O.}~\bibnamefont
  {Mendoza-Yero}}, \bibinfo {author} {\bibfnamefont {G.}~\bibnamefont
  {M{\'{i}}nguez-Vega}},\ and\ \bibinfo {author} {\bibfnamefont
  {J.}~\bibnamefont {Lancis}},\ }\bibfield  {title} {\bibinfo {title}
  {{Encoding complex fields by using a phase-only optical element}},\ }\href
  {https://doi.org/10.1364/OL.39.001740} {\bibfield  {journal} {\bibinfo
  {journal} {Opt. Lett.}\ }\textbf {\bibinfo {volume} {39}},\ \bibinfo {pages}
  {1740} (\bibinfo {year} {2014})}\BibitemShut {NoStop}%
\bibitem [{\citenamefont {{Luis Mart{\'{i}}nez Fuentes}}\ and\ \citenamefont
  {Moreno}(2018)}]{LuisMartinezFuentes2018}%
  \BibitemOpen
  \bibfield  {author} {\bibinfo {author} {\bibfnamefont {J.}~\bibnamefont
  {{Luis Mart{\'{i}}nez Fuentes}}}\ and\ \bibinfo {author} {\bibfnamefont
  {I.}~\bibnamefont {Moreno}},\ }\bibfield  {title} {\bibinfo {title} {{Random
  technique to encode complex valued holograms with on axis reconstruction onto
  phase-only displays}},\ }\href {https://doi.org/10.1364/OE.26.005875}
  {\bibfield  {journal} {\bibinfo  {journal} {Opt. Express}\ }\textbf {\bibinfo
  {volume} {26}},\ \bibinfo {pages} {5875} (\bibinfo {year}
  {2018})}\BibitemShut {NoStop}%
\bibitem [{\citenamefont {Wang}\ and\ \citenamefont
  {Piestun}(2020)}]{Wang2020}%
  \BibitemOpen
  \bibfield  {author} {\bibinfo {author} {\bibfnamefont {H.}~\bibnamefont
  {Wang}}\ and\ \bibinfo {author} {\bibfnamefont {R.}~\bibnamefont {Piestun}},\
  }\bibfield  {title} {\bibinfo {title} {{Azimuthal multiplexing 3D diffractive
  optics}},\ }\href {https://doi.org/10.1038/s41598-020-63075-8} {\bibfield
  {journal} {\bibinfo  {journal} {Sci. Rep.}\ }\textbf {\bibinfo {volume}
  {10}},\ \bibinfo {pages} {6438} (\bibinfo {year} {2020})}\BibitemShut
  {NoStop}%
\bibitem [{\citenamefont {Davis}\ \emph {et~al.}(2021)\citenamefont {Davis},
  \citenamefont {Wolfe}, \citenamefont {Moreno},\ and\ \citenamefont
  {Cottrell}}]{Davis2021}%
  \BibitemOpen
  \bibfield  {author} {\bibinfo {author} {\bibfnamefont {J.~A.}\ \bibnamefont
  {Davis}}, \bibinfo {author} {\bibfnamefont {E.~D.}\ \bibnamefont {Wolfe}},
  \bibinfo {author} {\bibfnamefont {I.}~\bibnamefont {Moreno}},\ and\ \bibinfo
  {author} {\bibfnamefont {D.~M.}\ \bibnamefont {Cottrell}},\ }\bibfield
  {title} {\bibinfo {title} {{Encoding complex amplitude information onto
  phase-only diffractive optical elements using binary phase Nyquist
  gratings}},\ }\href {https://doi.org/10.1364/osac.418578} {\bibfield
  {journal} {\bibinfo  {journal} {OSA Continuum}\ }\textbf {\bibinfo {volume}
  {4}},\ \bibinfo {pages} {896} (\bibinfo {year} {2021})}\BibitemShut {NoStop}%
\bibitem [{\citenamefont {Mansour}\ and\ \citenamefont
  {Papazoglou}(2018)}]{Mansour2018}%
  \BibitemOpen
  \bibfield  {author} {\bibinfo {author} {\bibfnamefont {D.}~\bibnamefont
  {Mansour}}\ and\ \bibinfo {author} {\bibfnamefont {D.~G.}\ \bibnamefont
  {Papazoglou}},\ }\bibfield  {title} {\bibinfo {title} {{Tailoring the focal
  region of abruptly autofocusing and autodefocusing ring-Airy beams}},\ }\href
  {https://doi.org/10.1364/OSAC.1.000104} {\bibfield  {journal} {\bibinfo
  {journal} {OSA Continuum}\ }\textbf {\bibinfo {volume} {1}},\ \bibinfo
  {pages} {104} (\bibinfo {year} {2018})}\BibitemShut {NoStop}%
\bibitem [{\citenamefont {Cottrell}\ \emph {et~al.}(2009)\citenamefont
  {Cottrell}, \citenamefont {Davis},\ and\ \citenamefont
  {Hazard}}]{Cottrell2009}%
  \BibitemOpen
  \bibfield  {author} {\bibinfo {author} {\bibfnamefont {D.~M.}\ \bibnamefont
  {Cottrell}}, \bibinfo {author} {\bibfnamefont {J.~A.}\ \bibnamefont
  {Davis}},\ and\ \bibinfo {author} {\bibfnamefont {T.~M.}\ \bibnamefont
  {Hazard}},\ }\bibfield  {title} {\bibinfo {title} {{Direct generation of
  accelerating Airy beams using a 3/2 phase-only pattern}},\ }\href
  {https://doi.org/10.1364/ol.34.002634} {\bibfield  {journal} {\bibinfo
  {journal} {Optics Letters}\ }\textbf {\bibinfo {volume} {34}},\ \bibinfo
  {pages} {2634} (\bibinfo {year} {2009})}\BibitemShut {NoStop}%
\end{thebibliography}%

\end{document}